%% file: main.tex
\documentclass[preprint,journal]{vgtc}            

\ieeedoi{xx.xxxx/TVCG.201x.xxxxxxx}

\title{DataWink: Reusing and Adapting SVG-based Visualization Examples with Large Multimodal Models}

\author{%
  \authororcid{Liwenhan Xie}{0000-0002-2601-6313},
  \authororcid{Yanna Lin}{0000-0003-3730-0827},
  \authororcid{Can Liu}{0000-0002-1175-0734},
  \authororcid{Huamin Qu}{0000-0002-3344-9694},
 and  \authororcid{Xinhuan Shu}{0000-0002-9736-4454}
}

\authorfooter{
  \item
  	Liwenhan Xie, Yanna Lin, and Huamin Qu are with the Hong Kong University of Science and Technology.
  	E-mail: \{liwenhan.xie, ylindg\}@connect.ust.hk, huamin@ust.hk,
  \item
  	Can Liu is with Nanyang Technological University.
  	E-mail: can.liu.1996@gmail.com,

  \item Xinhuan Shu is with Newcastle University.
  	E-mail: xinhuan.shu@newcastle.ac.uk. She is the corresponding author.
}

\abstract{%
  \input{meta/abs}
}


\teaser{
  \centering
  \includegraphics[width=\linewidth, alt={Applications of the proposed method.}]{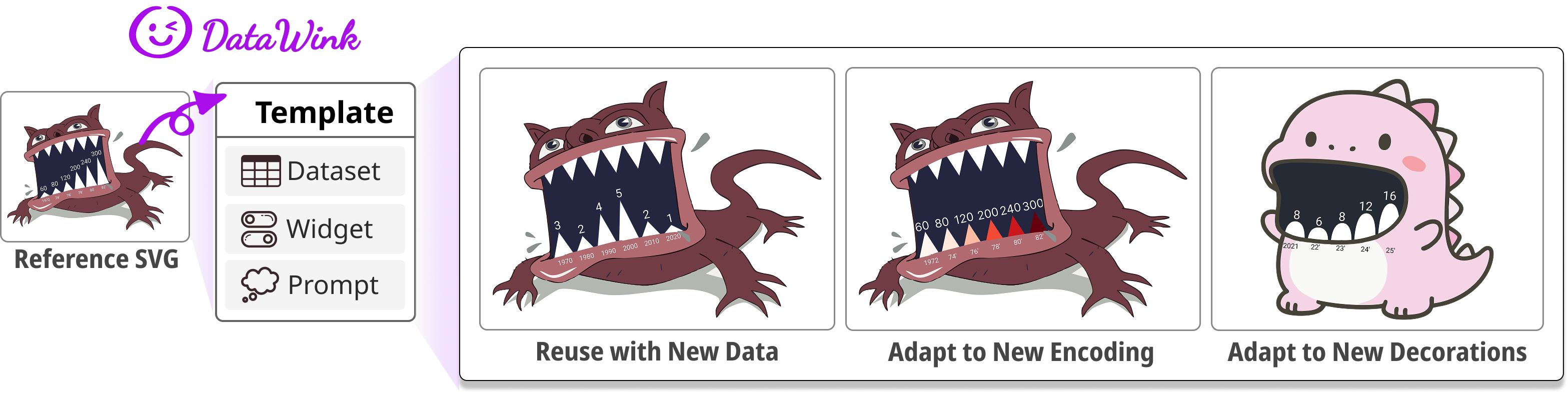}
  \caption{%
  	We introduce \tool, an interactive by-example authoring tool for reusing and adapting SVG-based visualizations.
    Powered by large multimodal models (LMMs), \tool automatically transforms the reference into an extensible template supporting dynamic controls through text and direct manipulation. With \tool, creators may update data or visual mapping scheme while maintaining original visual styles, and fine-tune the original image with an enriched SVG representation.
  }
  \label{fig:teaser}
}

\graphicspath{{figs/}{figures/}{pictures/}{images/}{./}} 

\usepackage{tabu}                      
\usepackage{booktabs}                  
\usepackage{lipsum}                    
\usepackage{mwe}                       

\usepackage{mathptmx}                  

\input{meta/packages}

\begin{document}


\firstsection{Introduction}

\maketitle

\input{src/01-introduction}

\input{src/02-review}
\input{src/03-decompose}

\input{src/04-template}

\input{src/05-interface}

\input{src/06-walkthrough}

\input{src/07-implementation}
\input{src/08-gallery}
\input{src/09-eval}
\input{src/10-dicussion}

\input{src/11-conclusion}
\input{meta/supp}
\input{meta/ack}
\newpage
\bibliographystyle{abbrv-doi-hyperref}
\balance
\bibliography{reference}

\end{document}

%% file: meta/abs.tex
Creating aesthetically pleasing data visualizations remains challenging for users without design expertise or familiarity with visualization tools.
To address this gap, we present DataWink, a system that enables users to create custom visualizations by adapting high-quality examples.
Our approach combines large multimodal models (LMMs) to extract data encoding from existing SVG-based visualization examples, featuring an intermediate representation of visualizations that bridges primitive SVG and visualization programs.
Users may express adaptation goals to a conversational agent and control the visual appearance through widgets generated on demand.
With an interactive interface, users can modify both data mappings and visual design elements while maintaining the original visualization's aesthetic quality.
To evaluate DataWink, we conduct a user study (N=12) with replication and free-form exploration tasks.
As a result, DataWink is recognized for its learnability and effectiveness in personalized authoring tasks.
Our results demonstrate the potential of example-driven approaches for democratizing visualization creation.

%% file: meta/packages.tex
\graphicspath{{figs/}{figures/}{pictures/}{images/}{./}} 

\usepackage{tabu}                      
\usepackage{booktabs}                  
\usepackage{lipsum}                    
\usepackage{mwe}                       

\usepackage{mathptmx}                  
\usepackage[table,xcdraw,svgnames]{xcolor}
\usepackage{booktabs} 
\usepackage{soul} 
\usepackage[normalem]{ulem}
\usepackage{xspace}
\usepackage[pagebackref, bookmarks]{hyperref} 
\usepackage{listings}
\usepackage{wrapfig}
\usepackage{graphicx,calc}
\usepackage[most]{tcolorbox}  
\usepackage{setspace}  
\usepackage{listings}
\usepackage{epigraph}
\usepackage{balance}
\usepackage{sparklines}
\usepackage[most]{tcolorbox}

\definecolor{hahablue}{HTML}{4d6ab8}
\definecolor{aliceblue}{rgb}{0.94,0.97,1.0}
\definecolor{lavendermist}{rgb}{0.9, 0.9, 0.98}
\definecolor{azure(web)(azuremist)}{rgb}{0.94, 1.0, 1.0}
\definecolor{mytextcolor}{HTML}{005682}
\definecolor{mycolor}{rgb}{0.90,0.90,0.90}
\definecolor{myblue}{HTML}{0c096b}
\definecolor{mygreen}{HTML}{406300}
\definecolor{myred}{HTML}{440072}
\definecolor{mybluecolor}{HTML}{f4fcff}
\definecolor{mypurplecolor}{HTML}{f9eeed}
\definecolor{mygreencolor}{HTML}{f4fcf5}
\definecolor{navy}{RGB}{52, 85, 139}
\definecolor{darkpurpletext}{HTML}{651491}
\tcbset{on line, boxsep=.8pt, left=0pt,right=0pt,top=0pt,bottom=0pt, colframe=white,colback=mycolor,code={\singlespacing}}

\definecolor{codebg}{rgb}{0.96,0.96,0.96} 
\definecolor{codeborder}{rgb}{0.85,0.85,0.85} 

\lstdefinestyle{prettyinline}{
  backgroundcolor=\color{codebg},     
  basicstyle=\ttfamily\small,         
  breakatwhitespace=false,            
  breaklines=true,                    
  captionpos=b,                       
  commentstyle=\color{gray},          
  keywordstyle=\color{blue},          
  stringstyle=\color{teal},           
  numbers=none,                       
  numberstyle=\tiny\color{codeborder},
  rulecolor=\color{codeborder},       
  frame=single,                       
  framexleftmargin=2pt,               
  framexrightmargin=2pt,              
  framextopmargin=2pt,                
  framexbottommargin=2pt,             
  framesep=2pt,                       
  xleftmargin=3pt,                    
  xrightmargin=3pt,                   
  aboveskip=5pt,                      
  belowskip=3pt,                      
  showstringspaces=false,             
  upquote=true,                       
}

\AtBeginDocument{

}

\newcommand{\eg}{e.g.,}
\newcommand{\ie}{i.e.,}
\newcommand{\etal}{et al.}
\newcommand{\etc}{etc.}

\def\tool{DataWink\xspace}
\def\user{Lily\xspace}

\def\repo{\url{https://github.com/shellywhen/DataWink}}

\newcommand{\rev}[1]{ {\color{black}{#1}}}

\usepackage{enumitem}  

\setlist[itemize,1]{label=$\bullet$, itemsep=0pt, leftmargin=1.2em, topsep=0pt, nosep} 
\setlist[itemize,2]{label=$\bullet$, nosep, leftmargin=1.5em} 
\setlist[itemize,3]{label=$\circ$, nosep, leftmargin=1.5em} 
\setlist[enumerate]{parsep=0pt, partopsep=0pt, itemsep=0pt, topsep=0pt}

\definecolor{lightyellow}{HTML}{FFF9C4}
\sethlcolor{lightyellow}
\newcommand{\finding}[1]{#1}

\newlength\myheight
\newlength\mydepth
\settototalheight\myheight{Xygp}
\newcommand{\iq}[1]{\textcolor{black}{\it``#1''}}  
\newcommand{\qb}[1]{\textcolor{black}{\small (#1)}} 
\newcommand{\rp}[1]{\textcolor{black}{\small #1}}  

\newcommand{\vmark}[1]{%
    \begingroup
  \setlength{\fboxsep}{0pt}%
  \colorbox{gray!5}{%
    \raisebox{0pt}[0.6\height][0.3\height]{%
      \hspace{0.1em}
      \textcolor{teal}{\texttt{<#1>}}%
      \hspace{0.1em}
    }%
  }%
  \endgroup
}

\newcommand{\prop}[1]{%
    \begingroup
  \setlength{\fboxsep}{0pt}%
  \colorbox{gray!5}{%
    \raisebox{0pt}[0.6\height][0.3\height]{%
      \hspace{0.1em}
      \textcolor{darkgray}{\texttt{#1}}%
      \hspace{0.1em}
    }%
  }%
  \endgroup
}

\newtcolorbox[]{mybox}{
    colback=white,   
    colframe=black,  
    boxrule=1pt,   
    sharp corners,    
    boxsep=4pt,
}

%% file: src/01-introduction.tex
\epigraph{{\it Immature poets imitate; mature poets steal.}}{T. S. Eliot}
\vspace{-1em}
\noindent Data visualization has evolved beyond mere functional representation to encompass highly engaging and aesthetically pleasing forms of expression, as exemplified by the {\it Dear Data} project~\cite{lupi2016Dear}.
Research shows that aesthetically pleasing visualizations enhance viewer engagement, memorability, and information retention~\cite{harrison2015infographic,borkin2016beyond}.
However, creating such compelling visual representations remains a significant barrier for general users who lack formal design training or technical knowledge of visualization tools.
While professional designers can masterfully balance visual elements to create engaging data representations~\cite{bigelow2016iterating, parsons2022Understanding, cairo2023art}, novice users often struggle to achieve comparable results using existing tools, creating a notable divide between professional-grade visualizations and those created for personal use.

To mitigate the challenge of conceiving vivid graphical design in data visualization, a line of research \cite{ying2022MetaGlyph,zhang2020DataQuilt, coelho2020infomages, snyder2025challenges} has investigated re-purposing or re-designing a given example instead of constructing from scratch.
For instance, MetaGlyph~\cite{ying2022MetaGlyph} supports binding data to a layered vector art.
Some works attempt to infer data mapping from examples, such as timelines~\cite{zhu2019towards}, bar charts~\cite{cui2021mixed}, and pictograms~\cite{shi2022supporting}, then generate a template for modification.
However, these approaches may fail in creative visualization design, where complex relationships between graphical elements are introduced beyond standardized charts or common mark types.
For instance, in \cref{fig:interface}-A, the shadows are parallelograms rather than rectangles. Their lengths must be consistent with the parents (the windows) and maintain a cohesive connection with their parent objects at the point of contact.
Moreover, prior works provide few editorial layers beyond primitive attributes, restricting the expressiveness of the final output and interaction effectiveness.

In light of recent advances in generative AI (GenAI), we are interested in addressing this gap in recovering data encoding visualizations intertwining with graphical embellishments.
Recent evidence shows that large multimodal models (LMMs)\footnote{While ``Multimodal Large Language Model (MLLM)'' is often used for language-first multimodal models, we adopt ``LMM'' in this paper for simplicity.} may generate source code for complex visual outputs~\cite{ shi2024ChartMimic} and support contextual modification~\cite{kim2022stylette, vaithilingam2024dynavis, goswami2025plotedit}.
However, under the context of personal visualization reuse, two key challenges persist.
First, the relationships of visual elements may not follow a standard hierarchy or pattern, making it challenging to predefine rules to recover encoding schemes.
Second, decorative elements may serve multiple purposes (\eg~purely aesthetic versus functional like gridlines), making them integral to a visualization. Misidentifying dependencies can lead to corrupted designs when reused with new data.

We investigate an automatic pipeline powered by LMMs that transforms graphical primitives into reusable templates for user-driven adaptation.
As a preliminary step, we focus on visualization examples represented in Scalable Vector Graphics (SVG), a prevalent format in web-based visualizations and a structured representation that preserves the hierarchical relationships between visual elements, making it more amenable to analysis than pixel-based images.
Our method attends to both the graphical integrity and design integrity, shaping the visual style of a visualization, ensuring dynamic adaptation to the user's new data and context.
On the one hand, the graphical reference is transformed into a layered representation for synthesizing generator programs of data-driven elements;
On the other hand, the low-level program can be transformed into a dynamic parametric template for creators to refine and redesign.
We also design a tool featuring dynamic template generation and bespoke widget synthesis from user commands, which aims to lower the authoring barrier.
We demonstrate the practicalness and expressiveness of our approach in a self-curated gallery by reusing and adapting various public visualization designs.
A user study (N=12) evaluated the usefulness and effectiveness of \tool, where participants generally recognized the learnability of the system and enjoyed the by-example authoring workflow in personal contexts.

In general, our study contributes to the following aspects.

\begin{itemize}[leftmargin=*]
      \item An automatic pipeline that (a) transforms low-level graphical primitives into semantically rich representations for program synthesis, and (b) scaffolds these representations with reusable, parameterized templates for user-driven customization.
    \item \tool, a by-example authoring tool that encapsulates the proposed pipeline. Using the reconstructed template inferred from a reference, users may apply personalized data and adaptations.
      \item A user study (N=12) evaluating the usefulness and effectiveness of the tool for example reuse and adaptation. We further discuss design implications for creative visualization authoring workflows.
\end{itemize}

%% file: src/02-review.tex
\section{Background \& Related Work}

\subsection{Reusing Examples in Visualization Creation}
Conventional visualization authoring tools (\eg~\cite{kim2016data, liu2018dataillustrator,  ren2019Charticulator, offenwanger2024datagarden}) often assume a clear vision in the creators and require precise specification of the visualization structure upfront, which may pose initial challenges to design ideation and implementation~\cite{satyanarayan2019critical}.
In practice, visualization creators may draw inspiration from examples and perform design remixing~\cite{bigelow2014Reflections, bako2022Understanding, baigelenov2025how}, a common approach in web design~\cite{warner2023interactive, kumar2011bricolage} and graphics design~\cite{lu2024Misty}.
As such, there have been ongoing efforts to support users to borrow, adapt, and refine existing design elements, fostering a more explorative and iterative workflow.
Relevant works typically facilitate reusing visualization specifications~\cite{mcnutt2021integrated, harper2018Converting, cui2021mixed, tsandilas2020structgraphics}, repurposing design assets like readily-crafted graphical components~\cite{zhang2020DataQuilt, ying2022MetaGlyph}, extracting graphical primitives like color palettes~\cite{yuan2021deep, shi2022supporting}, and transferring abstract styles like layout patterns~\cite{song2024gvvst} and motion dynamics~\cite{xie2023wakey}.

We are interested in reusing the entire visual structure in creative, aesthetically pleasing visualization designs.
With a similar goal, prior works have investigated reusing different types of visually-enriched visualizations, including timelines~\cite{zhu2019towards}, bar charts~\cite{cui2021mixed}, proportion-based visualizations~\cite{qian2020retrieve}, and pictorial visualizations~\cite{shi2022supporting}.
For instance, to reuse timelines, Zhu-Tian~\etal~\cite{zhu2019towards} proposed a pipeline that first classifies design options, then parses the content and roles of visual elements, and finally generates an extensible template, where users fill in a \texttt{JSON} specification for reusing the design.
Our work complements existing works by incorporating the relationship between data-driven elements and external visual design.
Instead of deconstructing the example into discrete design options that may fail creative designs, we maintain the raw SVG-based representation and infer graphical constraints in template generation.
To effectively surface the extracted data encoding scheme and graphical constraints, we design an interactive template following a data-agnostic design workflow~\cite{tsandilas2020structgraphics}.

\subsection{Reverse Engineering \& Visual Structure Extraction}
Reverse engineering of visualization aims to recover data values and corresponding visual encoding schemes from a low-level graphical representation of data visualizations, which remains a fundamental task for digitalization, automatic analysis, and redesign of data visualizations~\cite{wu2021ai4vis}.
A majority of works focusing on raster images~\cite{savva2011revision, poco2017ReverseEngineering,luo2021chartocr} or vector graphics~\cite{masson2023ChartDetective, chen2024Mystique, jung2017chartsense, mendez2016iVoLVER}, follow a similar approach to classify charts into specific types, identify data-driven graphical marks, and recover the data.
Some recent works explored end-to-end inference, leveraging visual reasoning capabilities in large multimodal models~\cite{shi2024ChartMimic} or training models on an enhanced dataset incorporating visual structures~\cite{dou2024Hierarchically}.
However, existing techniques cannot be directly applied to our scope, as we focus on creative visualization designs with rich graphical decorations and free-form layouts, which may violate assumptions in predefined heuristics or yield out-of-distribution errors.

Another research stream seeks to computationally interpret patterns or structures for graphics or UI design, where
a core task is to identify the roles of visual constituents~\cite{madan2021parsing, zhang2023editing, liu2018learning} and model their relationships~\cite{kembhavi2016diagram, shin2021multi, shi2023understanding}.
Specific to data-driven graphics, Lu~\etal~\cite{lu2020vif} developed a method to construct the semantic structure that links graphical elements to convey information, namely visual information flow.
Zhou~\etal~\cite{zhou2024DataPictorial} proposed a pipeline that converts raster images into SVG objects with attribute values binding to data, assuming the original dataset is available.
Continuing prior efforts, we propose a method that addresses reverse engineering of visually rich visualizations featuring nuanced associations between graphical elements.

Most relevant to our work, Harper and Agrawala~\cite{harper2014deconstructing, harper2018Converting} converted D3-based visualizations into structured templates, leveraging embedded information.
Qi~\etal~\cite{qi2025sketch2data} supports data recovery from expressive hand-drawn infographics using a user-defined parametric glyph template.
In comparison, our approach directly infers data values underlying the example and enables user adaptation to decorative elements, moving beyond visual encodings.
Moreover, our user interface features dynamic templates and text-based editing, which further reduces the learnability.

\subsection{LMM-powered Visualization Authoring Support}
Recent advances in LMMs have sparked research interest in enhancing authoring tools with AI-powered capabilities for more expressive visual outputs and efficient user workflows~\cite{di2023doom,he2025Leveraging}.

On the one hand, LMMs facilitate generating compelling designs for infographics~\cite{zhou2024epigraphics, xiao2024typedance}, pictorial visualizations~\cite{xiao2023spark}, and data analogies~\cite{chen2024beyond}.
Using image generation models, some studies proposed novel pipelines for end-to-end generation, combating their inherent limitations to preserve structural relationships in data encoding~\cite{xiao2023spark, chen2024beyond}.
Some follow a bottom-up approach to recommend design components~\cite{xiao2024typedance, zhou2024epigraphics}.
For instance, Zhou~\etal~\cite{zhou2024epigraphics} proposed to guide asset curation from an epigraph and supports between-asset fine-tuning.

 \begin{figure*}[t]
    \centering
    \includegraphics[width=0.9\linewidth]{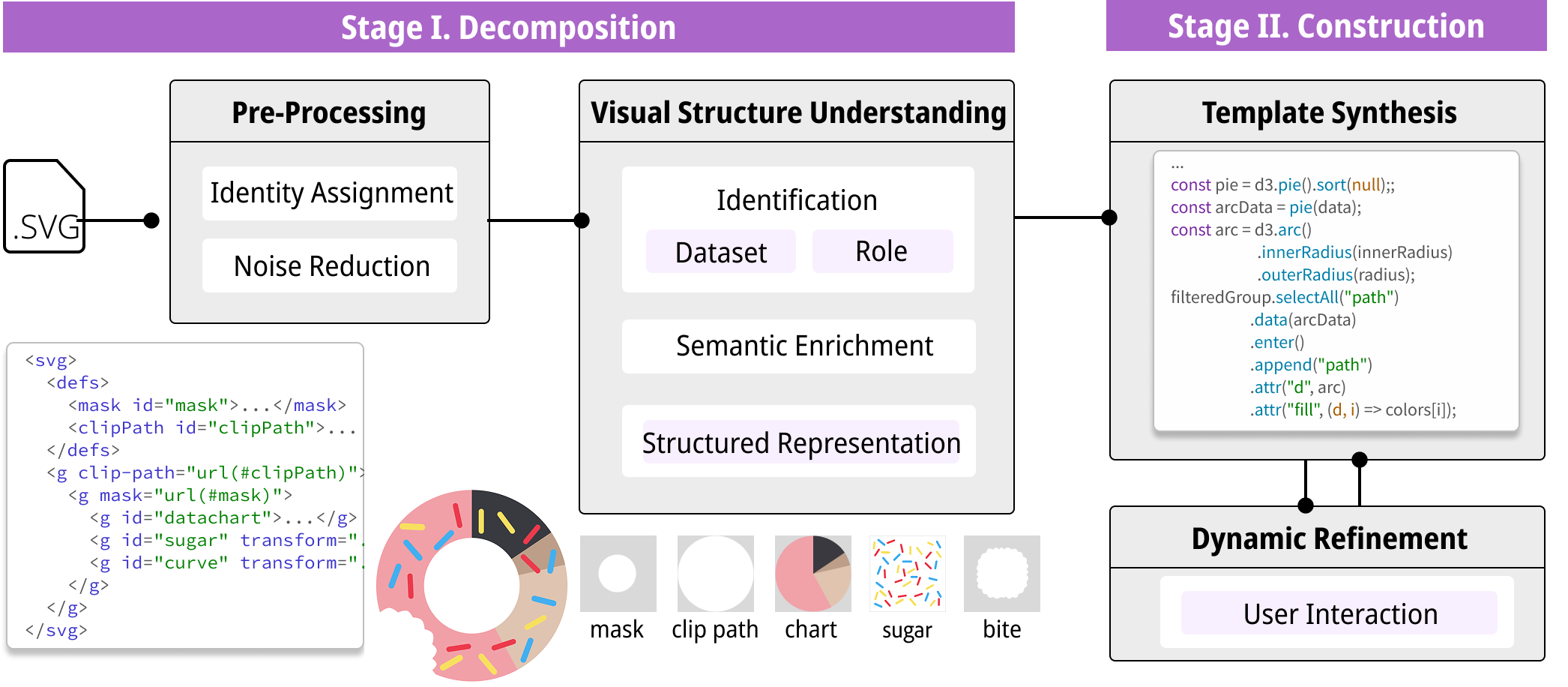}
    \caption{An overview of the pipeline. (I) Decomposing an example visualization into an intermediate representation. (II) Constructing reusable templates as a combination of code snippets.}
    \label{fig:pipeline}
\end{figure*}

On the other hand, LMMs excel at coding and natural language understanding.
Many studies have investigated natural-language guided generation of data visualizations, \eg~\cite{tian2024chartgpt}.
For more flexible user control beyond textual prompts, some works explore alternative interaction paradigms to express user intents, \eg~\cite{xie2024waitgpt, vaithilingam2024dynavis, shen2024data, wang2025DataFormulator2, cheng2024biscuit}.
Xie~\etal~\cite{xie2024waitgpt} introduced an interactive graph-based representation to facilitate steering LLM-generated code for data processing and encoding.
Recognizing the need to dynamically adjust visual encoding schemes in iterative design, Vaithilingam~\etal~\cite{vaithilingam2024dynavis} proposed to generate control widgets based on editing commands.
More recently, Wang~\etal~\cite{wang2025DataFormulator2} blended graphical user interfaces and natural language inputs to enable direct selection and iterative refinement.

This paper investigates an understudied LMM-powered authoring workflow to reuse SVG-based visualizations, taking advantage of the capacity in visual reasoning and structured document generation in LMMs~\cite{liu2024logomotion}.
We leverage an abstraction based on SVG representation to facilitate reverse engineering by LMMs, and design a user-friendly interface to scaffold low-level data encoding schemes.
Amidst the wave to harness LMMs in visualization tasks, we contribute a novice-friendly method to reuse and adapt the creative design of visualizations.

%% file: src/03-decompose.tex
\section{From Example to Reusable Template}
We propose an LMM-powered pipeline to synthesize a reusable template from an SVG-based visualization featuring advanced visual designs. Here, we elaborate on each module and technical details.


\subsection{Pipeline Overview}
As shown in \cref{fig:pipeline}, our pipeline consists of two phases: (I) decomposition--decomposing the visual structure into a semi-structured representation from a visualization reference, and (II) construction--constructing an extensible template from the extracted visual structure.
The input of the pipeline is a visualization reference formatted in SVG.
The output is a visualization template in the form of a functional generator accepting various parameters, such as datasets, color palettes, and chart-specific configurations like bar chart gaps.

In the decomposition phase, there are two modules: SVG pre-processing and visual structure understanding.
The first module simplifies the SVG file by removing unnecessary details, making it easier for the LMM to process.
The second module restructures the SVG specification into semantically rich layers, extracts critical visual information for reverse engineering, and infers the concrete visual encoding scheme and the precise data used in the reference.
In the reconstruction phase, based on the cleaned reference SVG and the extracted visual encoding scheme, an LMM is prompted to synthesize a D$^3$-based program~\cite{bostock2011d3} for generating a raw SVG string of the data-driven layers, which serves as the template.
With the template, users may update the data and visualization specifications dynamically.

Our approach focuses on SVG-based visualizations with rigid, data-driven elements.
Free-form hand-drawn sketches, such as in DataInk~\cite{xia2018dataink} and DataGarden~\cite{offenwanger2024datagarden}, are beyond the scope of this work.

\subsection{Intermediate Representation of Visualization}
\label{sec:representation}
While SVG preserves low-level graphical details in a visualization, it loses a direct association with the data to be updated and challenges effective reuse, let alone adaptation.
Therefore, a key step in our approach is to represent the reference visualization in a layered abstraction, which categorizes visual components of different roles and aggregates their attribute values and spatial relationships.
SVG-based visualization reference includes four layer types: data-driven layers, text layers, decorative layers, and configuration layers.

\textbf{Data-driven Layers} are essentially chart components, which can be classified into data marks, axes, and legends, according to their roles~\cite{liu2025manipulable}.
To support reuse, a core challenge is to identify the data mapping scheme underlying visual elements in the data-driven layers. 
As such, we define an abstraction shown in the card below.

\noindent\begin{mybox}

  \textit{\textbf{Global Properties}}
    \begin{itemize}
    \item Coordinate := (Cartesian | Polar | Customized)
        \item Origin: (x\_pos, y\_pos)
        \item Canvas bounding box: (x\_pos, y\_pos, width, height)
        \item Prototype: (Bar chart | Scatterplot | Line | Area | Radar | ...) 
    \end{itemize}
\vspace{\baselineskip}
\textit{\textbf{Layer-wise Properties}}
    \begin{itemize}
        \item Visual marks
        \begin{itemize}
            \item Mark type := [(deformed path | trend path | atomic shapes)]
            \item Data-encoded attributes
            \item Fixed attributes
        \end{itemize}
        \item Axis
            \begin{itemize}
                \item Grid line
                \item Labels
            \end{itemize}
        \item Legend
            \begin{itemize}
                \item Position: (x\_pos, y\_pos)
                \item Size: (width, height)
                \item Visual channel groups
            \end{itemize}
    \end{itemize}

\end{mybox}
\vspace{-0.5\baselineskip}

\textbf{Text Layers} are text elements without a determinate relationship to the visualization, such as titles, captions, and annotations.
Different from data-driven text like tick labels, the content is at a higher abstraction level and thus requires user steering.

\textbf{Decorative Layers} are standalone image assets that serve for visual decoration without data-driven elements.
A common example is a background or foreground image.

\textbf{Configuration layers} are necessary dependency items for the SVG file not being directly rendered in the visual output.
These include the \vmark{style} element defining the cascading style sheets (CSS) rules that apply to the SVG, the \vmark{defs} element defining reusable elements (\eg~ patterns, filters, gradients, or symbols), the \vmark{script} element defining interaction event handlers, animation, and others.

\subsection{Phase I. Decomposing SVG-based Visualizations}
In Phase I, the underlying data and data encoding scheme are extracted from the input SVG-based visualizations.

\subsubsection{SVG Pre-processing}
To fit the LMM's context window and facilitate visual reasoning tasks, the SVG is initially preprocessed to remove redundant information.

\textit{Identity Assignment.}
At first, we create a unique identity number for each visual element embedded in the SVG string as a new attribute.
This serves later reconstruction with a key mapping mechanism.

\textit{Noise Reduction.}
We then apply several heuristics to simplify the SVG string.
An intuition behind these heuristics is to provide a reference value other than precise for LMMs without depriving it of attention to details.
For instance, we directly empty the \prop{innerHTML} of descriptive elements as this hardly affects the final visual outputs, such as the \vmark{style}, \vmark{filter} elements, \etc~
In addition, We simplify complex shapes by reducing the number of control points while preserving their overall appearance.
Their control points or numerical attributes are simplified by retaining two decimal places.

\subsubsection{Visual Structure Understanding}
With the simplified SVG, we constructed a three-step LMM chain for extracting the intermediate representation.
Each step's output serves as the input for the subsequent step.
The chain also incorporates a raster image of the visualization for enhanced context.
The image is empirically resized proportionally to a max width of 400 pixels for text legibility while minimizing tokens.
To enhance accuracy, we embed an example of a bar chart in the prompt and illustrate the target output.

\textit{Role Identification \& Data Extraction.}
 An LMM with a long context window is prompted to group relevant visual elements into one of the four predefined abstract layers for different roles in a visually rich visualization, \ie~data-driven layer, text layer, decorative layer, and configuration layer, and then reorganize the inner components formatted as an SVG string.
 As introducing a new \vmark{g} element (a tag for containers commonly used for grouping elements) may break the original styles, we specifically require grouping these elements using self-developed XML markers invalid in the SVG standard for later decoupling.
 In addition, the LMM should also synthesize a dataset matching the values presented on the rendered visualization.
 Note that we merge the two tasks here to simplify the pipeline.
Data extraction is independent of role identification.

\textit{Semantic Enrichment.}
To connect the raw SVG string with its rendered visual contents better, we prompt another LMM to write brief descriptions in natural language for element groups and embed them into group attributes, following Liu~\etal~\cite{liu2024logomotion}.
For data-driven layers, the LMM should classify element groups into visual marks, axes, and legends, according to the representation aforementioned.

\textit{Intermediate Representation Generation.}
Finally, with a clean, role-specific, and semantically rich SVG output, the LMM is explicitly prompted to refine and finalize the structure, elaborated in \autoref{sec:representation}.

\vspace{.5\baselineskip}
\noindent This chaining strategy is adopted to address the inherent limitations of LMMs in directly parsing and reasoning over complex structures within an SVG file.
By decomposing the task into smaller, well-defined steps, the approach reduces ``cognitive overhead'' for LMMs, enabling it to incrementally build understanding through progressively enriched intermediate representations.

%% file: src/04-template.tex
\subsection{Phase II. Constructing Reusable Templates}
\rev{Phase II synthesizes a template that combines the marked-up SVG and the encoding scheme into a program, whose arguments correspond to stylization parameters, such as the layout configuration, color palettes, \etc, which can be dynamically updated in the user's conversational interactions and link back to the GUI widgets for fine-grain refinement.}

\subsubsection{Template Synthesis}
\rev{Based on the structural representations, this step leverages an LMM to synthesize a corresponding template, where the goal is to capture the essence of the reference design and support personalized reusing and adaptation.
To this end, we pose three constraints to the target program.
First, the input of the program includes i) the marked-up SVG reference, ii) the data to visualize, and iii) other visual parameters.
Second, the program generates data-driven elements and inserts them into the marked-up SVG or manipulates the SVG.
This preserves the original graphical design and its relationship with data-driven elements (\eg~the bite shape and mask in \cref{fig:pipeline}).
Third, all variables should be accessed through a list of objects.
To accommodate basic needs in reusing a visualization, such as data mapping and layout adjustment, this list of objects contains some fixed variables, including the chart width, chart height, origin positions, and data attributes for axes.

In the LMM prompt, we clarify that the goal is to synthesize a program and associated variable list meeting the requirements aforementioned while replicating the reference as accurately as possible, using the parsed original data.
We also adopt a one-shot learning strategy with one example to guide manipulating the marked-up SVG reference.
Our implementation leverages \texttt{D3.js}~\cite{bostock2011d3} as the base library due to its modular and declarative nature, which aligns well with LMMs’ reasoning capabilities.
This allows the generated program to bind data to visual elements and define intricate relationships efficiently.
}


\begin{figure*}[t]
    \centering
    \includegraphics[width=\linewidth]{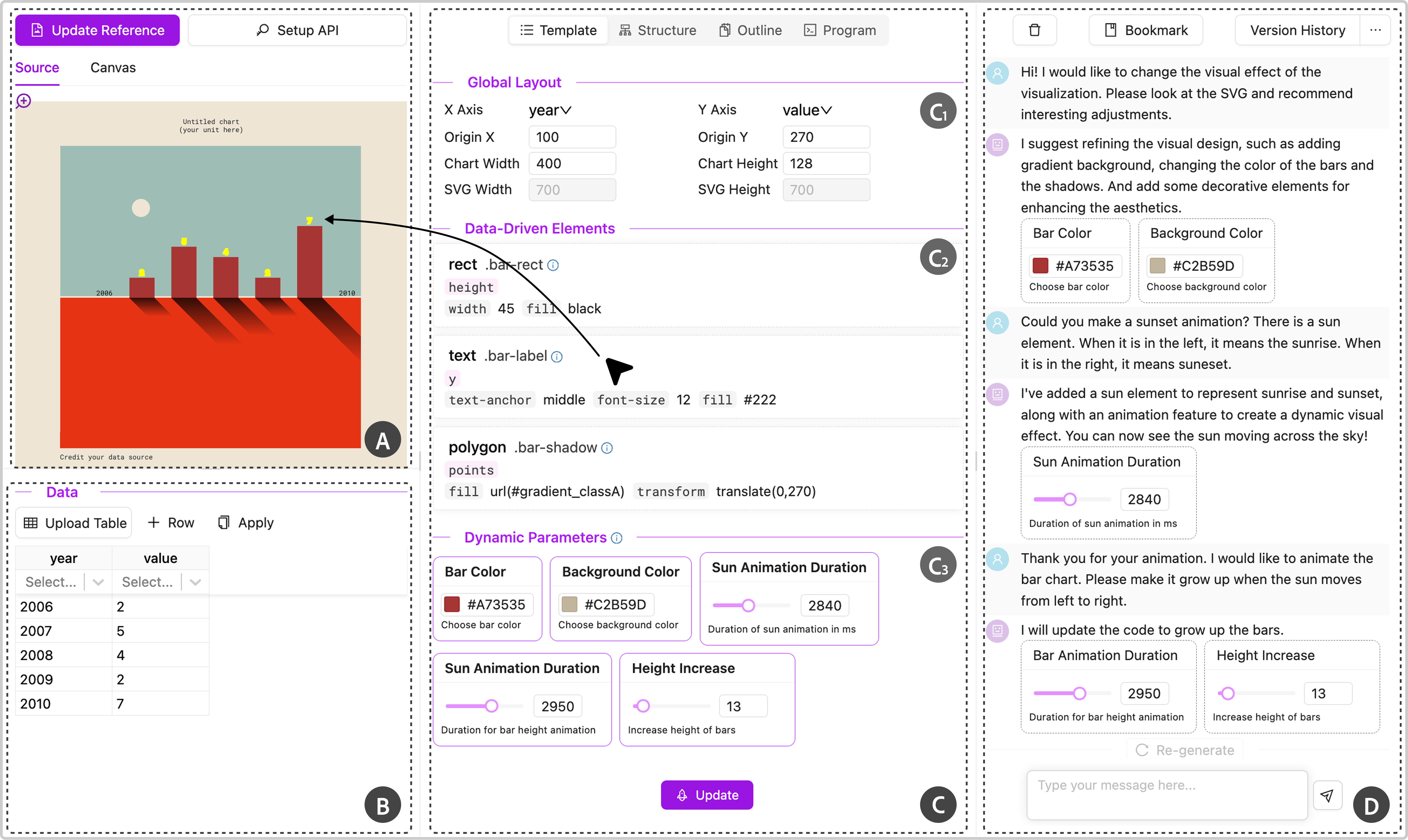}
    \caption{A screenshot of \tool, the proof-of-concept authoring interface for reusing and adapting SVG-based examples. (A) Canvas panel--Users can upload a reference, browse different bookmarked visualization designs, and edit the visual elements in the generated layer.
    (B) Data panel--The inferred data is displayed in an interactive table. Users may upload their data and select corresponding columns to fit in the inferred schema.
    (C) Template panel--Users can change the global layout (C$_1$), inspect the identified data-driven layers (C$_2$), and update the visual encoding scheme (C$_3$). (D) Through a conversational interface, users may refine the visual design through widgets generated on demand.
    }
    \label{fig:interface}
\end{figure*}
\subsubsection{Dynamic Refinement}
\label{sec:refinement}


\rev{The visualization template can be further updated through its input configurations.
Given a user request for adaptation, an LMM is prompted to decide on appropriate updates to the program, which may include adding new parameters, updating existing ones, modifying internal processing logics, or revising the marked-up SVG structure.
For new parameters, the LMM should choose from a list of provided widgets (slider, color picker, input, \etc) and specify their properties.
Despite alternatives in updating the program, we prioritize minimal changes.

For instance, when the user requests a thinner bar, the LMM may introduce a parameter \texttt{bar\_width} to the program, and specify its property being a numerical value suitable for a slider with associated attributes like title, max value, min value, and recommend a default value smaller than the previous hard-coded constant.
The user can then drag the rendered slider widget if not satisfied with the default.
}

%% file: src/05-interface.tex
\section{\tool: Interface}
\label{sec:interface}
With the proposed pipeline for decomposing and templating an SVG-based visualization, we design \tool\footnote{The name ``\tool'' implies creating beautiful visualizations adaptive to personal design requirements or data context in a wink.}, which aims to support creating bespoke visualizations by reusing high-quality designs.

\subsection{Design Goals}
We distill the challenges underlying the envisioned workflow from the literature and formulate our design goals. 

\begin{enumerate}[leftmargin=2.75em]
\renewcommand{\labelenumi}{\textbf{DG\theenumi.}}
\renewcommand{\labelenumii}{ DG\theenumi.\arabic{enumii}}

\item Minimize user interaction in adapting new data. 
Typically, reusing requires aligning new data with the visual structure of an existing visualization. 
To streamline this process, we aim to minimize user interactions by providing an interactive configuration panel to fit user data into the parsed data schema.

\item Surface data encoding schemes with dynamic templates~\cite{vaithilingam2024dynavis, {zhu2019towards}}.
With a reference SVG, a key foundation is to create an extensible and dynamic template that abstract graphical elements into meaningful data-driven components.
It can further be easily rendered with different datasets and adjusted by parameters. 

\item Support contextual adaptation and redesign from high-level textual prompts. 
Adapting and redesigning complex designs often require diverse UI operations and iterative refinements.
Instead of directly modifying the visualization output, our approach translates textual prompts into structured UI components that scaffold the editing process and enable control.

\item Support adaptation and redesign from direct manipulation.
Instead of solely depending on textual prompts, certain refinements, such as adjusting layouts and resizing elements, are more intuitively and precisely handled through direct manipulation with the visualization itself. 

\end{enumerate}

\subsection{System Overview}
\Cref{fig:interface} shows a screenshot of \tool.
It comprises a canvas panel, a data panel, a template panel surfacing different levels of representations for the working visualization, and a chat panel to communicate with an LMM-powered assistant. 
Each panel is resizable and collapsible.

\vspace{0.5\baselineskip}
\noindent\textbf{Canvas Panel.}  
The canvas panel, located on the left, displays both the original SVG reference and the user-generated visualization (\cref{fig:interface}-A).
Users can switch between these views using a tab menu.
Depending on the underlying program, the SVG-based visualizations may support interactivity and animations.
The canvas is visually linked to the template, with corresponding elements highlighted in yellow for better context.
Additionally, users can directly resize and reposition the entire chart or decorative elements directly on the canvas (DG4).

\vspace{0.5\baselineskip}
\noindent\textbf{Data Panel.} The data panel has an interactive table displaying the working data (\cref{fig:interface}-B).
By default, the inferred data for the reference visualization is shown here.
Users may update the table through direct manipulation or by uploading a CSV file with the same data types (DG1).
To accommodate new data with a different naming system, the user may specify the correspondence through a drop-down menu.

\vspace{0.5\baselineskip}
\noindent\textbf{Template Panel.}
The template panel in the middle shows configurable attributes for the extracted template, including the fixed visual mapping configuration, \ie~global layout (\cref{fig:interface}-C1), identified data-driven elements (\cref{fig:interface}-C2), and a list of dynamic widgets generated through conversation to control visual attributes (\cref{fig:interface}-C3).
Users may inspect the extracted details about the identified data-driven elements by hovering over cards that specify each data-driven element.
They can also directly manipulate the parameters (DG3) to see the visual output.
When a user interacts with the template, the visualization will be automatically regenerated with the new parameters.

To make the internal mechanism transparent, we provide alternative views for the underlying template.
The second view titled ``Structure'' displays the marked-up SVG, which embeds information about the roles of elements and the extracted semantics in an interactive tree.
With the hidden SVG identifiers, it visually links to the working visualization on the canvas, 
providing a complementary view of the generated result.
Through the tree structure, advanced users can change the hierarchy, properties, and groupings (DG4).
The third view titled ``Outline'' displays the intermediate representation for the reference.
The fourth view titled ``Program'' displays the underlying D$^3$ code to manipulate the SVG reference and current parameter settings.

\begin{figure*}[t]
    \centering
    \includegraphics[width=\linewidth]{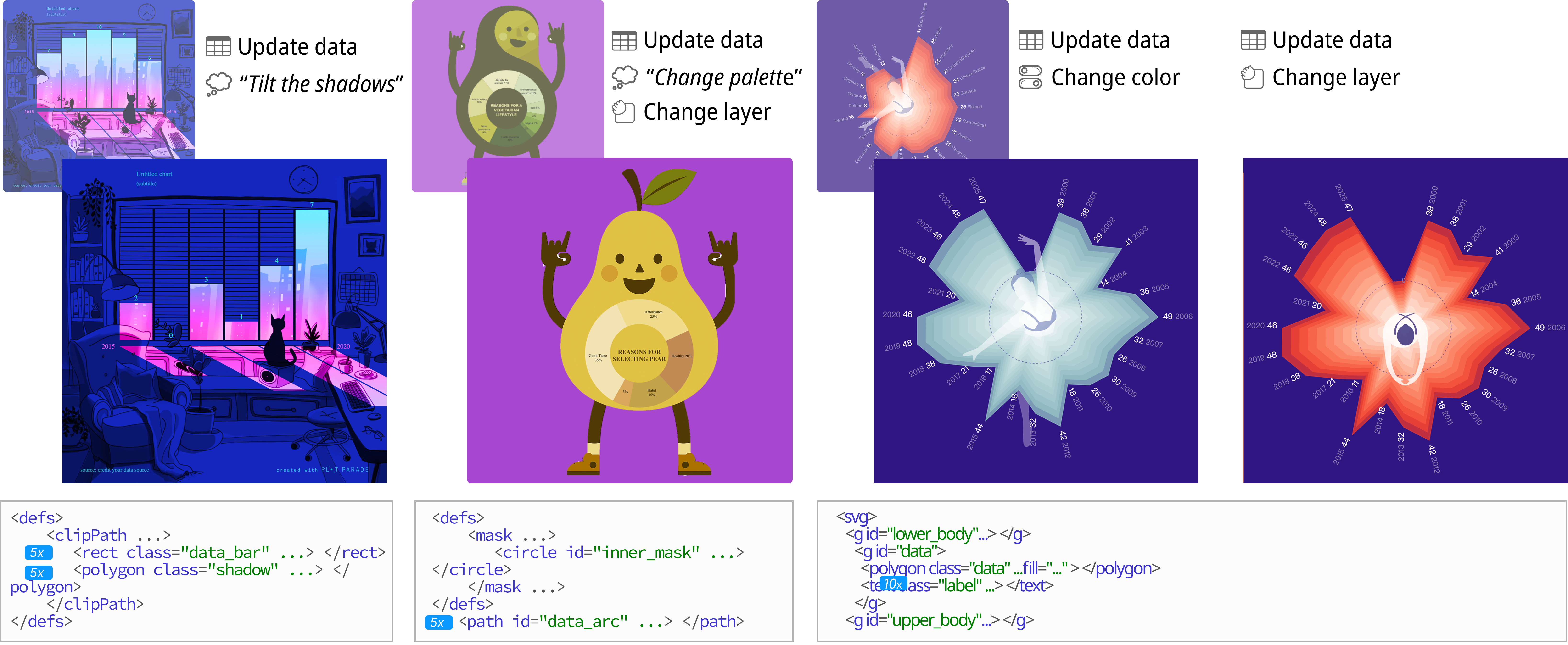}
    \caption{A gallery of visualizations made with \tool reusing reference SVGs (top left) and adapting the visual design. \rev{Sources from left to right: Krisztina Szűcs's \textit{Window} ©, Amber Zuñiga's \textit{What You Should Know About Vegetarianism} ©, and Valentina D'Efilippo's \textit{Gender Wage Gap Viz} ©.}}
    \label{fig:gallery}
\end{figure*}
\vspace{0.5\baselineskip}
\noindent\textbf{Chat Panel.}
\rev{The chat panel supports natural language-based interactions (\cref{fig:interface}-D), where users may talk to an LMM-powered chat assistant embedding the parsed visualization template context.
}
When users pose instructions for updating the visualization, the canvas panel will be updated in response to the user's requirements (DG3), and corresponding UI widgets (\eg~value sliders, color pickers, and text inputs) will be generated and inserted into the current chat thread and the dynamic widget panels within the template panel (DG4).
Users may leverage the UI widget to adjust granular parameters to control the visual output.
Each text-based interaction corresponds to a system checkpoint logging the current state.
Users may bookmark certain checkpoints and resume previous states during design iteration.

Internally, the LMM assistant is prompted with the user query, contextual information including the raster image of the working canvas, and template details (\ie~the function, marked-up SVG, and default parameters).
\rev{If the user requests adaptation, it generates an updated program and associated parameters, as elaborated in \autoref{sec:refinement}.}
When new parameters are introduced, their specifications will be parsed and linked to a live widget.
\rev{Otherwise, the LMM will reply in text only.}

%% file: src/06-walkthrough.tex
\subsection{Usage Scenario}
Here, we walk through a hypothetical usage scenario to demonstrate how the authoring tool may facilitate reusing a visualization reference.
The final output is illustrated in \cref{fig:teaser} (Right).

\user is a volunteer in an animal rescue center and an amateur data enthusiast.
For an upcoming community event, she plans to create a poster featuring data facts about animals.
She was interested in adapting the famous {\it Monstrous Costs Chart} created by Nigel Holmes, as she would like to resonate with the residents through the fun figure.

\textbf{Data Adaptation.} \user uploads an SVG-based replication of the visualization she found on the Internet.
After several minutes, a template is generated and the data in the original chart has been recovered, including a column named ``year'' and a column named ``value''.
\user then inputs corresponding data to get an updated chart.

\textbf{Encoding Adaptation.} Viewing from the template panel, \user sees that the data is bound to deformed paths, \ie~the sharp triangular shape.
She feels this design is too aggressive for her intended cute theme.
She then chats with the system chatbot: ``Replace the triangular teeth with soft, rounded shapes''.
The system interprets her request and updates the path elements accordingly, resulting in a gentler and more approachable teeth shape.

\textbf{Design Refinement.}
Next, \user decides to change the monster into a more pleasant figure that better suits the new rounded teeth.
She exports the generated SVG from the structure panel.
With SVG editing tools, she removes the background layer annotated as ``background'', adds a pink dinosaur illustration image generated by GPT,
and adjusts the layout and scale of the inserted background to integrate it with the rest of the visualization.

%% file: src/07-implementation.tex
\subsection{Implementation}

\tool is a web-based app implemented in the React framework based on TypeScript.
Both the template generation pipeline and the conversational assistant are powered by OpenAI's API.
Specifically, we adopted the \texttt{GPT-4o-mini} and the \texttt{GPT-4o-128k} model (only in the pre-processing step of Phase I), applying a suite of prompt templates for structured outputs with minimum prompt engineering effort.
The source code and prompt templates are available at \repo.

%% file: src/08-gallery.tex
\section{Gallery}
\rev{\Cref{fig:gallery} showcases three examples using \tool to reuse and adapt real-life instances that blend basic charts with graphical design, covering bar chart, radar chart, and pie chart.}
\begin{figure*}
    \centering
    \includegraphics[width=\linewidth]{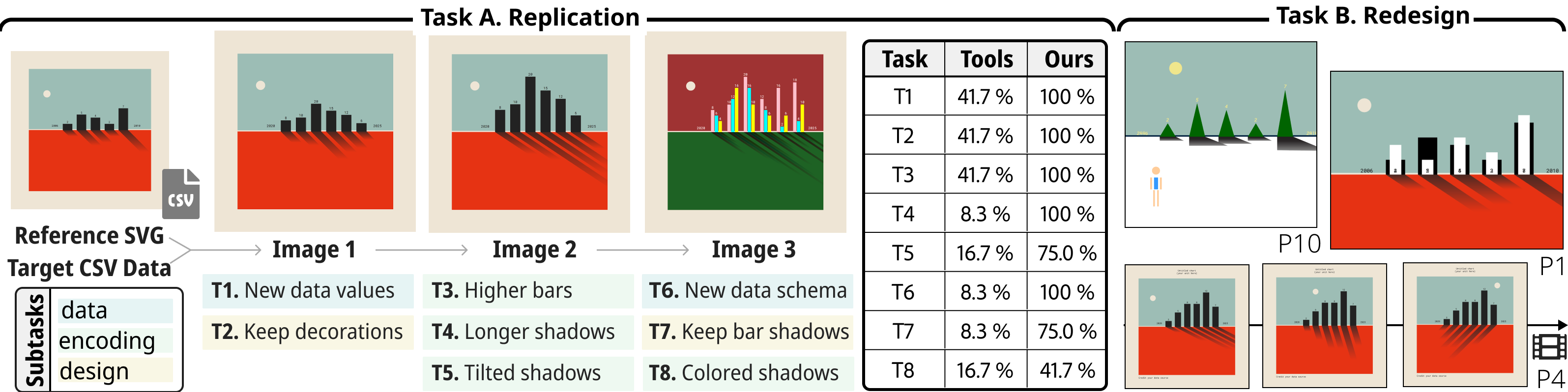}
    \caption{An overview of tasks in the user evaluation. In Task A, participants are provided with a reference SVG, a dataset, and three images to replicate, with 8 subtasks on data- (blue), encoding- (green), and style-based (yellow) adaptation. In Task B, participants redesign the visualization freely, expressing their creative agency. Left: Task A materials and completion criteria. Middle: The completion rate (\%) for the individual replication task. Right: Exemplar outputs in Task B.
    P10 transformed the scene into a snowy cypress forest and added a human figure. P1 updated the data schema and tried with nested bar charts. P4 created an animation with a shadow-sun interplay mimicking daylight changes.
    }
    \label{fig:task}
\end{figure*}

%% file: src/09-eval.tex
\section{User Evaluation}
\label{sec:eval}
To evaluate the authoring experience with \tool, we conducted a user study (N=12).
We aim to address the following questions.
\begin{enumerate}[leftmargin=25pt]
    \renewcommand{\labelenumi}{\textbf{RQ\theenumi.}}
    \item Does the templating approach in \tool effectively facilitate reusing a bespoke, SVG-based data visualization? (DG1 \& 2)
    \item Does the dynamic widget generation mechanism in \tool effectively aid in adapting a visualization example? (DG3 \& 4)
    \item How does \tool align with users’ natural workflows?

\end{enumerate}

\subsection{Protocol}
We conducted a user study using a within-subjects design. 
Each session was conducted one-on-one and consisted of a replication task with two conditions, followed by an open-ended redesign task using \tool only. 
Participants were compensated at approximately \$14.50 per hour, which aligns with the local median wage.

\vspace{0.5\baselineskip}
\noindent\textbf{User Tasks.}
In the user study, there are two tasks: a replication task\footnote{The reference SVG sources are from \url{https://plotparade.com/}, a signature work by Krisztina Szűcs.} and a redesign task, namely Task A and Task B.
\Cref{fig:task} illustrates an overview of the subtasks and result summary.

\begin{itemize}
    \item Task A: Design Replication.
    We require the participants to create a visualization based on a given data formatted in CSV, an SVG-based reference visualization with graphical decorations, and three raster images showing the target output after reusing and adapting the reference.
    This task aims to mimic the process of realizing a visualization design from a mental model, utilizing provided references as concrete guides for precise replication.
    
    \item Task B: Redesign.
   Participants are allowed to personalize and adapt the visualization from Task A, pushing beyond the initial replication.
   This task simulates a scenario to extend and transform an external design, allowing creative explorations.
\end{itemize}

\vspace{0.5\baselineskip}
\noindent\textbf{Baseline \& Apparatus.} \rev{As there was no available software that directly addressed reusing non-standard charts, we allowed participants to mix using commercial tools they were familiar with in the baseline condition.
These tools represented the status quo that users would otherwise rely on to complete similar tasks.
Available tools included conversational AI assistants (\eg~OpenAI ChatGPT and Google Gemini), vector graphics editors (\eg~Adobe Illustrator and Figma), and code editors enabling AI features (\eg~Cursor and Microsoft VSCode).}
\tool was deployed as a web-based application.
Participants completed the tasks on their laptops, either in person or via Zoom.

\vspace{0.5\baselineskip}
\noindent\textbf{Procedure.}
Before the study, participants signed the consent form and provided basic information.
Besides, we provided dedicated accounts for the software used in the baseline conditions if needed.
The formal study lasted about 96 minutes \rp{(SD=17, Range 65--125)}\footnote{The large variance was due to the early exit of tasks or various model response times dependent on participants' locations.}.

\begin{itemize}
    \item Briefing (15 min): The facilitator briefly introduced the purpose of the project, and then walked the participants through the \tool using a toy example of a donut chart.
    \item Task Execution ($50$ min): Participants completed Task A and B sequentially.
    In Task A, there was a 15-minute time limit per condition.
    A counterbalanced design was employed: half completed the baseline condition first, while the others started with \tool.
    In Task B, there was a 5-minute ideation phase to encourage creative exploration, followed by a 15-minute redesign activity. 
    \item Questionnaire and Interview (15 min): Participants filled out a survey hosted on Google Forms and joined a post-study interview.
\end{itemize}

The time constraints for individual stages were informed by two pilot studies.
During task execution, participants were encouraged to think aloud and explain their intentions.
The facilitator would pause the study and assist if unexpected errors occurred.

\vspace{0.5\baselineskip}
\noindent\textbf{Threats to Validity.}
This study involved a small sample size (N=12), which may limit the generalizability of the findings. Additionally, the use of commercial tools as baselines might not reflect the full range of tools available for visualization reuse and redesign.

\subsection{Participants}
We recruited participants through social media and word-of-mouth, where we specifically sought individuals interested in creating bespoke communicative visualizations.
To ensure a diverse range of perspectives, we included users with varying skill levels in design and visualization authoring.
This approach aimed to better understand the challenges and strategies involved in using the proposed workflow.

The study involved 12 participants (6 females, 6 males) aged 23--30 years \rp{(M=26.17, SD=2.12)} with diverse backgrounds in HCI, visualization, media arts, and AI.
All participants shared a common cultural and linguistic background.
On average, participants had 3.41 years of experience in visualization authoring \rp{(SD=2.15, range: 0-8 years)}, 4.83 years in graphic design \rp{(SD=2.95, range: 1-11 years)}, and 3.08 years in visualization programming \rp{(SD=2.31, range: 0–8 years)}.
Nine participants (75\%) reported taking inspiration from or directly reusing reference visualizations.
On a 5-point Likert scale, participants rated their familiarity with visualization authoring as moderately high \begin{sparkline}{5} \setlength\sparkbottomlinethickness{0.2pt} \definecolor{sparkbottomlinecolor}{gray}{0.3} \sparkspike .15 0 \sparkspike .317 .16 \sparkspike .483 .5 \sparkspike .65 1 \sparkspike .817 .16 \sparkbottomline[0.9] \end{sparkline} \rp{(M=3.75, SD=0.87)}.
In the following, they are denoted as P1--P12, ranked by their years of experience with data visualization authoring.

\subsection{Measurements \& Analysis}
We embedded logging functionalities in \tool.
Each session was screen-recorded and audio-recorded.
We employed both quantitative and qualitative measures in the analysis. 
Participants completed a questionnaire to assess usability, effectiveness, and satisfaction centered around our research question using 7-point Likert-scale questions.
\rev{The usability questions were borrowed from the System Usability Scales (SUS)~\cite{lewis2018system}.}
Additionally, the post-study interviews were transcribed and coded thematically to identify common patterns, challenges, and user strategies.
Key themes included usability, workflow integration, and perceived pros and cons of \tool.

\subsection{Quantitative Results}
\label{sec:results}

\textbf{Task A: Task Completion.} The table in \cref{fig:task} compares the task completion rates between \tool and commercial tools familiar to the participants, where \finding{participants in the \tool{} condition achieved a significantly higher performance}.
Specifically, with \tool, all participants (N=12) could finish data adaptation tasks, maintain original visual decorations, and perform global chart adjustment.
Despite this, some participants (N=4) failed to control the data-driven shadows and ran out of time before synthesizing accurate gradients.
With the baseline, more participants (N$\ge$9) encountered challenges in creating and manipulating shadows based on new data.

\vspace{0.5\baselineskip}
\noindent\textbf{Task B: Redesign Activities.} Participants have explored diverse redesigns, such as altering metaphorical data representations (N=4), updating chart types (N=3), adding animation (N=3), and enabling interactions (N=2).
Some redesign activities are data-independent, such as modifying the color scheme (N=2) or adding graphical elements (N=4).
On average, participants engaged in 10.58 rounds of chat \rp{(SD=3.87)} and utilized 8.50 widgets \rp{(SD=3.68)} throughout the process.
The screenshot in \cref{fig:interface} shows how P4 rapidly tests out ideas. 

\vspace{0.5\baselineskip}
\noindent\textbf{Questionnaire: Self-Reported Ratings.}
\Cref{fig:rating} shows responses to the post-study questionnaire.
\finding{Participants found it easier to complete Task A in \tool \rp{(M=6.17, SD=.72)} than in tools that they are familiar with \rp{(M=2.50, SD=1.57)}}.
And they were more satisfied with the quality of replicated visualizations using \tool \rp{(M=5.91, SD=.67)} than others \rp{(M=2.58, SD=1.88)}.
Their perceptions were significantly different under a pairwise t-test (p<.0001 for both questions).
In general, \finding{\tool received predominantly positive ratings across ease of use, customization, and integration into existing workflows.}

\subsection{Findings}
\label{sec:findings}

\begin{figure}[t]
    \centering
    \includegraphics[width=\linewidth]{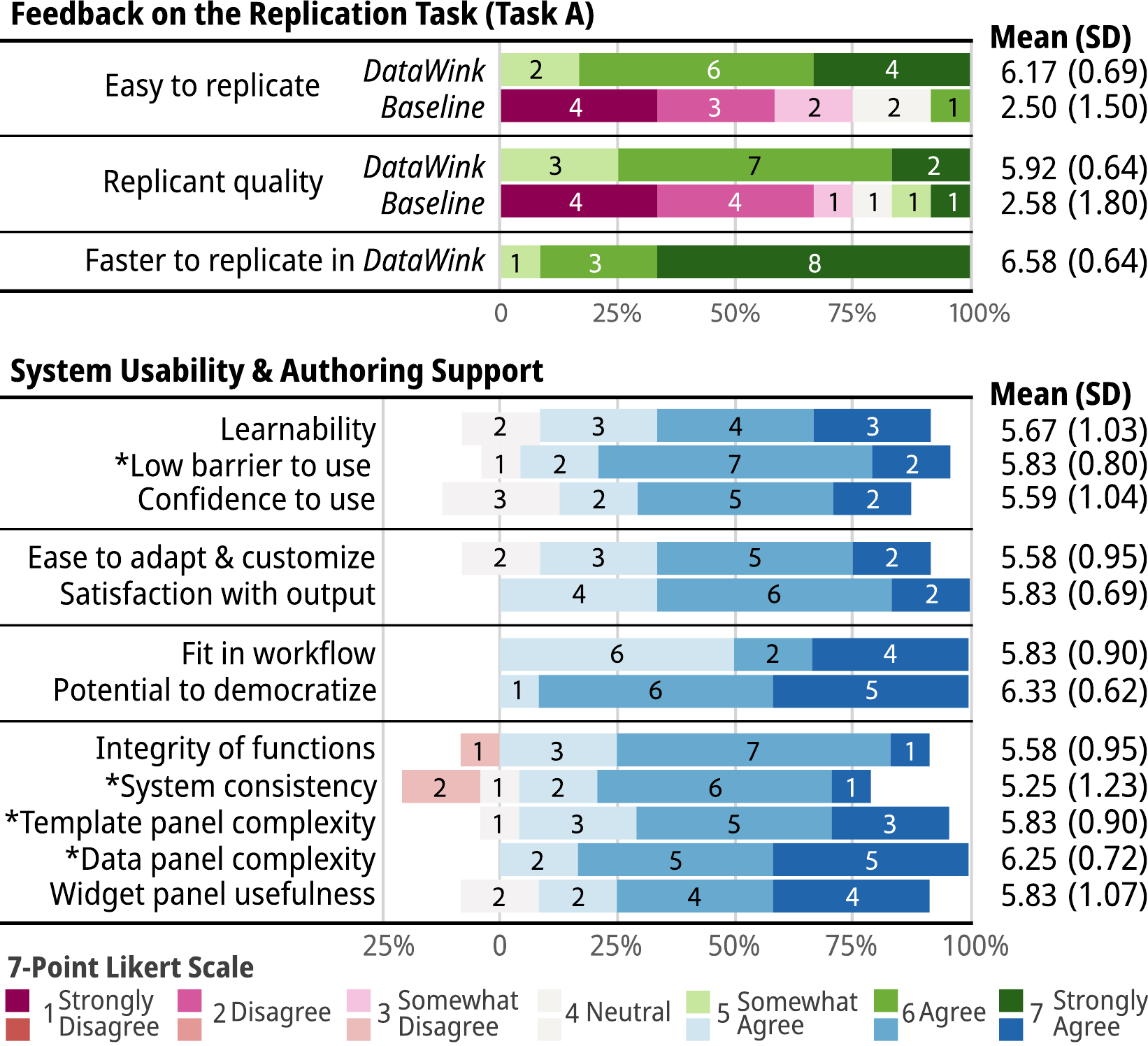}
    \caption{Distribution of user ratings based on a 7-point Likert scale. 1: Highly disagree; 4: Neutral. 7: Highly agree. Top: Comparison between the baseline and \tool. Bottom: Evaluations on the usefulness and effectiveness of \tool.  Note that some questions were originally posed in the reverse form; for reporting clarity, these items were flipped to imply stronger agreement with a large value (annotated with *).}
    \label{fig:rating}
\end{figure}

\noindent\textbf{RQ1. Effectiveness in Reusing a Reference.}
In Task A T1-2, T6-7, most participants ($\ge$75\%) can adapt the reference visualization to a given data while maintaining the original styles, outperforming the baseline condition (\cref{fig:task}).
Additionally, participants self-rated it easier to replicate a visualization with \tool, and expressed greater satisfaction with the output (\cref{fig:rating}-A).
They strongly agreed that \tool supports faster reuse compared with the baseline.

{\large $\diamond$} \uline{Challenges in reusing references with \rev{existing} tools.}
In the baseline condition, participants employed a variety of tools and strategies, each with its own limitations, detailed as follows.

For SVG editing tools (P2-6, P9, P11), it remained tedious to curate data-driven elements.
Four participants struggled with creating shadowed parallelograms, as these shapes are not predefined.
Adjusting the bottom control points of grouped rectangles led to inconsistent shadow angles for different lengths.
\rev{To accomplish the task in time, some did not calculate the precise position but dragged the control point based on their rough visual estimation, leading to inaccurate visualizations.}
In addition, P3 pointed out scalability concerns in their workflow: \iq{I cannot deal with a larger dataset by tweaking element properties one by one.}
Besides, the unfamiliar SVG structure created barriers to effective reuse (N=3).
For instance, P9 struggled to select the bars due to an invisible mask at the upper layer that prevents direct manipulation.   

For LMM-powered workflows, participants requested direct generation of SVG strings (P7-8, P10) or raster images (P1).
\rev{They generally recognized the convenience of text-guided task execution.
}
Here, a significant challenge was to maintain graphical integrity (N=4), design integrity (N=2), ensure correct syntax (N=3), and generate a complete, displayable file (N=2).
For instance, P1 tried image generation with the latest \texttt{GPT\thinspace o3} and \texttt{GPT-4.5}, however, the data-driven marks hardly scaled based on the values.

Four participants requested an early exit due to frustration with the exhausting trial-and-error.
\rev{\iq{I know how to use Figma, but I had strong trust in GPT's capability and hope to finish the task fast. Now I feel deeply disappointed by so many random errors.} \qb{P7}}
\rev{Sensemaking of the SVG remains another challenge, as users need to articulate where to change and go into details.}
When an SVG is generated with errors with invalid values, users can be clueless about how to debug---\iq{The SVG string contains too many decimal spaces to browse through. I feel desperate not seeing the output!} \qb{P8}

{\large $\diamond$} \uline{Perceived strengths of \tool in reference reusing tasks.}
Many participants (N=9) shared excitement about being able to reuse a visualization reference by directly replacing the dataset (DG1), especially for those in the baseline-first group.
This \iq{eliminates the tedious procedure to establish concrete data mappings already there [in the reference]}\qb{P9}.
Most participants found the extracted template easy to understand (DG2, \rp{M=5.83, SD=.90}).
\finding{The interactive cards linking the parsed data-driven elements and the rendered elements on the canvas facilitate an intuitive understanding of the encoding scheme (N=6).}
P2 leveraged the identifiers and attribute names when constructing their prompts, highlighting its benefits for human-AI alignment: \iq{It [The template panel] dissects the data-driven elements from the visualization and provides meaningful explanations on how AI understands it.}

\vspace{0.5\baselineskip}
\noindent\textbf{RQ2. Effectiveness in Adapting a Visualization Template.} Participants largely agreed that it was easy to adapt and customize a reference using \tool \rp{(M=5.58, SD=0.95)}, and they were quite satisfied with the output \rp{(M=5.83, SD=0.69)}.


{\large $\diamond$} \uline{Benefits of language-guided widget synthesis} (DG3, DG4).
The participants found the widget panel useful \rp{(M=5.83, SD=1.07)}.
\finding{Complementing uncertainty in the text-based instruction, the UI widgets for newly introduced parameters facilitate convenient adjustments.}
\iq{Essentially, they [the synthesized widgets] helped me arrive at the right place without iterating with LLMs back-and-forth.}\qb{P12}
\finding{Another advantage is the support for rapid design exploration, where LMM can progressively unlock the expressivity of the underlying D$^3$-based visualization generator.}
\Cref{fig:task} showcases three examples of participants' outputs in Task B, demonstrating their innovation in reimagining existing works.
P2 appreciated the all-in-one workspace that allows refining data-driven elements and graphical embellishments through natural language---\iq{I can do it [editing] on an editing tool, but it is interesting to see how to control it [target element] through variables.}
In addition, P1 noted that a chatbot reduced the need to specify targets in design iterations: \iq{I feel more focused without naming the target I am referring to.}

{\large $\diamond$} \uline{Glitches in realizing authorial adaptation intents.}
Despite the convenience, we observed several glitches for participants to realize their authorial intents in natural language.
Firstly, articulating and communicating authorial intents requires a level of familiarity with terms.
In realizing the third image in Task A (see \cref{fig:task}), expert designers (P1-2, P4) effortlessly constructed prompts that elaborated on all requirements, such as explaining new data series, specifying web color names, and describing the nuances in shadow colors with a term like ``hint''.
In comparison, non-designers sometimes struggled to express their vision.
Secondly, many participants (N=7) felt constrained by the sole modality in text.
P6 desired object referral through direct manipulation---\iq{If only I can click to suggest which element I am talking about!}
\rev{Similarly, P7 envisioned a more natural communication paradigm using speech and gesture to indicate adaptation intents.}
Thirdly, similar to other LMM-powered tools, \tool might be prone to errors and hallucinations in the generated output (N=6), memory limits (N=3), and language ambiguity (N=2).
For instance, many participants (N=4) ran into an issue where \prop{\#sun} was selected despite the nonexistence of such an identifier.
Correspondingly, some participants (N=5) lacked confidence in using \tool.
\rev{Lastly, the LMM capability boundary can be unclear to the users.}
To tilt the shadows as in Task A T5, P11 wrote: \iq{Place the sun to the right}, assuming that the shadow would follow the sun internally.
However, identifying and maintaining such a physical relationship in SVG remained a challenge \cite{liu2024logomotion}.
\rev{In another example, P10 would like to switch the color scheme by saying ``{\it Change to the dark mode}'', which was not effectively addressed.}

\vspace{0.5\baselineskip}
\noindent\textbf{RQ3. Workflow Integration.}
As seen in \cref{fig:rating}, the overall feedback indicates that \tool offers a streamlined, flexible environment for reusing and adapting visualizations, with strong potential to democratize bespoke visualization authoring for newcomers and non-designers.

{\large $\diamond$} \uline{Abstraction matching in template onboarding and widget learning.}
Viewing \tool from a perspective in end-user programming, a key challenge is supporting \textit{abstraction matching}, connecting the users' high-level goals in reusing and adapting visualizations with the LMM-generated representations, \ie~the template and widgets.
We approached this through interactive scaffolds of the extracted data-driven elements and widget-controlled programs, allowing users to ``climb the abstraction ladder''.
As a result, participants generally believed others would easily learn to use the system \rp{(M=5.67, SD=1.03)}, reflecting the strong learnability of \tool.
While most participants focused solely on the template panel, some skillful users (N=4) also inspected the structure panel and the code panel.
P2 expressed their need for such an exposure to low-level details---\iq{I feel much more in control knowing the contextual information for AI.}
\rev{P5 shared a similar recognition: \iq{\rev{I may not look at them [intermediate representations] every time, but they should be there}}}.
The code panel also facilitated debugging when hallucinations arose (N=3).

{\large $\diamond$} \uline{General expectations for system improvement.}
Participants shared ideas for further strengthening \tool to empower visualization reuse and redesign.
P1 and P3 were visualization practitioners.
Both of them mentioned a need for random dataset generation covering the representative patterns for fast design evaluation.
P2 noted a practical need to revive an \iq{ugly chart} with a reference instead of starting design reuse from raw data:
\iq{A chart is more specific than a textual description of targets. This [Incorporating a chart input] may yield a more precise adaptation.}
\tool currently supports direct manipulation of the entire chart.
However, for more granular control, the parameters for data-driven elements can be inferred from demonstration---\iq{If I move one bar away, the gap should be adjusted automatically.}\qb{P6}.
P11 wished for a library of references to choose from and future support to combine multiple references into one.
P4 highlighted the playful and creative nature of the redesign process, suggesting embedding affective feedback to inspire and amplify user creativity.
\iq{If someone is cheering for my redesign and sharing their interpretations, I guess I will play with the system for a whole day.}

%% file: src/10-dicussion.tex
\section{Discussion}

\subsection{Limitation}

\noindent\textbf{Assumed accessibility of high-quality SVG-based examples.}
The input of \tool, \ie~SVG-based reference, represents only a small portion of the large volume of visualization images, which primarily consist of basic charts~\cite{chen2024visanatomy, zhang2023oldvisonline}.
Consequently, obtaining high-quality examples often relies on serendipity, which calls for incentives to foster creative design (\eg~\textit{Information is Beautiful} contest~\cite{iib}) and better infrastructure for reference-seeking (\eg~\cite{xiao2023wytiwyr}).  
Recent discussions on contemporary visualization design, such as Stefaner’s reflection on the ``crisis'' in innovation~\cite{stefaner2025crisis}, and Wu’s exploration of what stifles innovation~\cite{wu2025what}, point to growing homogeneity in visual styles as templated solutions dominate.  
Notwithstanding, we see our work as a positive effort against one-size-fits-all templates.  
By lowering the barriers to adapting and building upon the visual designs of references, we aim to stimulate creativity by standing on the giant's shoulders.  

\vspace{0.5\baselineskip}
\noindent\textbf{Challenges in reliability, efficiency, and adaptation with LMMs.}
Current LMMs have a limited understanding of SVG representations, likely due to the scarcity of training data~\cite{wu2024chat2svg, liu2024logomotion}.
Our pipeline introduces several descriptors (\ie~the intermediate representation, slot-based SVG, and generators for data-driven elements) to combat such a pitfall.
However, these additions can result in excessively long prompts, causing noticeable latency in LMM services (exceeding 20 seconds in some user study sessions), which negatively impacts user experience.
\rev{In addition, our method operates upon LMM chains, which are prone to propagated errors at early stages, such as misinterpreting SVG slots, leading to template generation failures.
This constrains its applicability to more advanced visualization types, such as glyph-based metaphoric visualizations.
}
Mitigation strategies, such as error detection, fallback mechanisms, or ensemble models, could improve robustness.
Last, text-based editing, while effective for certain design tasks, is not always suitable for graphical design workflows, which call for hybrid interaction paradigms, as discussed in \autoref{sec:findings}.

\vspace{0.5\baselineskip}
\noindent\textbf{Constrained degrees of freedom in creative design.}
Our pipeline adopts a top-down strategy to generate data-driven elements: global layout $\rightarrow$ axes scales $\rightarrow$ data-driven elements.
While this structured approach simplifies the design process, it limits creative flexibility, requiring creators to balance stability with the freedom to refine visualization specifications.
Adapting to different canvas sizes presents additional challenges. In the proposed workflow, changes in the canvas size require recalibration of global properties, such as layout, scales, and the positioning of data-driven elements.
These adjustments can disrupt the integrity of the design, making it difficult to maintain both aesthetic and functional consistency.
This challenge is especially critical in responsive design, where visualizations must dynamically adapt to varying screen sizes and aspect ratios~\cite{schottler2024constraint}.

\subsection{Future Works}
Our ultimate goal is to enable the flexible reuse of visualization references in design practices.
While this work focuses on SVG-based references due to technical feasibility, we envision broader opportunities and outline key research directions.

\vspace{0.5\baselineskip}
\noindent\textbf{Incorporating bitmap-based infographics in reverse engineering.}
By leveraging the standard DOM structure and native properties of SVG, we aim to decode the visual encoding scheme and allow designers to easily combine and modify existing visual components, fostering creativity and innovation.
With the emergence of generative AI, pictorial visualizations may blend with natural images~\cite{xiao2023spark, omer2024pictorialattributes} without a structured representation of data-driven variations, in contrast to the numeric attribute values inherent in a visual object in an SVG.
Future work will involve developing advanced algorithms that can intelligently analyze and deconstruct bitmap-based inputs, converting them into semantically rich layers with visual objects.
This will include exploring techniques for segmentation, role classification, and agentic workflow construction (\eg~\cite{ge2025autopresent}) for data-driven image asset generation.

\vspace{0.5\baselineskip}
\noindent\textbf{Relaxing intermediate representation into semi-structured relations.}
Another exciting avenue for future work is to explore an agentic framework that reverse engineers data mapping schemes through flexible visual abstractions, such as Bluefish~\cite{pollock2024bluefish} or hand-drawn ones~\cite{offenwanger2024datagarden, qi2025sketch2data}.
With practical ``tools'', LMMs can analyze, decompose, and reconstruct visualizations by bridging the gap between visual elements and underlying code representations.
To enhance this process, a self-evaluation scheme allows LLMs to assess their mapping quality and effectiveness instead of relying on one-shot trials.
For instance, when synthesizing a specific scale to map data values onto coordinate values, an LMM-powered agent may compare the output SVG and the original one to determine correctness.
This scheme empowers agents to critically examine outputs, align with established practices, and refine designs through quantitative metrics and user feedback.

\vspace{0.5\baselineskip}
\noindent\textbf{Towards a malleable interface for design remixing.}
We are also interested in the creative remixing process incorporating multiple references (\eg~\cite{shi2025brickify}).
We envision a malleable interface allowing users to reuse visual properties at different abstraction levels, from low-level style attributes like color palettes to high-level design patterns like layout dynamics. 
This flexibility would allow designers to leverage existing visualizations more effectively while maintaining creative control over their final designs in a progressive, iterative manner.
Following the grammar of graphics~\cite{wilkinson2011grammar}, visualization specifications like D$^3$~\cite{bostock2011d3} and Vega~\cite{satyanarayan2016vega} commonly ignore the dependency between graphical decorations and visualization design, assuming a sequential workflow to construct data-driven marks first and then add decorative layers or apply artistic deformations.
A more interesting direction is to explore representations that integrate graphical decorations and visualization design in a cohesive manner, allowing these elements to dynamically inform and influence each other to support a malleable interface.

%% file: src/11-conclusion.tex
\section{Conclusion}
This study explores supporting the reuse and adaptation of high-fidelity visualization designs, which involve complex graphical dependencies. To address challenges in preserving design intent and enabling efficient customization, we introduce an LMM-powered pipeline that converts a SVG-based visualization into a template. The template retain structural and stylistic elements while allowing dynamic adaptation to new datasets.
We develop \tool, an interactive authoring tool that integrates this pipeline with intuitive interfaces for by-example editing. Through a user evaluation, we demonstrate that \tool enables faster, more effective reuse and adaptation of visualizations compared to conventional tools while aligning with users’ creative workflows.

This work serves as a step toward empowering designers and non-expert users to leverage, adapt, and remix sophisticated visualizations, highlighting the transformative potential of LMMs in advancing the field of visualization design.
Future work will focus on broadening input modalities and expanding functionality to further empower users in creating and adapting visualizations.

%% file: meta/supp.tex







%% file: meta/ack.tex
\acknowledgments{
This work is partially supported by the Hong Kong RGC GRF Grant {\#16210722}. We thank the user study participants and appreciate the invaluable feedback from Prof. Fanny Chevalier.
}

%% file: main.bbl
\begin{thebibliography}{10}

\bibitem{baigelenov2025how}
A.~Baigelenov, P.~Shukla, and P.~Parsons.
\newblock How visualization designers perceive and use inspiration.
\newblock In {\em Proc. CHI},  article no. 1168,  13 pages. ACM, 2025. \href{https://doi.org/10.1145/3706598.3714191}
{doi: {{%
10\hspace{.1pt}\discretionary{.}{%
}{.}\hspace{.4pt}1145\discretionary{/}{%
}{/}3706598\hspace{.1pt}\discretionary{.}{%
}{.}\hspace{.4pt}3714191}}}


\bibitem{bako2022Understanding}
H.~K. Bako, X.~Liu, L.~Battle, and Z.~Liu.
\newblock Understanding how designers find and use data visualization examples.
\newblock {\em IEEE Trans. Vis. Comput. Graph.}, 29(1):1048--1058, 2022. \href{https://doi.org/10.1109/TVCG.2022.3209490}
{doi: {{%
10\hspace{.1pt}\discretionary{.}{%
}{.}\hspace{.4pt}1109\discretionary{/}{%
}{/}TVCG\hspace{.1pt}\discretionary{.}{%
}{.}\hspace{.4pt}2022\hspace{.1pt}\discretionary{.}{%
}{.}\hspace{.4pt}3209490}}}


\bibitem{bigelow2014Reflections}
A.~Bigelow, S.~Drucker, D.~Fisher, and M.~Meyer.
\newblock Reflections on how designers design with data.
\newblock In {\em Proc. AVI}, pp. 17--24. ACM, 2014. \href{https://doi.org/10.1145/2598153.2598175}
{doi: {{%
10\hspace{.1pt}\discretionary{.}{%
}{.}\hspace{.4pt}1145\discretionary{/}{%
}{/}2598153\hspace{.1pt}\discretionary{.}{%
}{.}\hspace{.4pt}2598175}}}


\bibitem{bigelow2016iterating}
A.~Bigelow, S.~Drucker, D.~Fisher, and M.~Meyer.
\newblock Iterating between tools to create and edit visualizations.
\newblock {\em IEEE Trans. Vis. Comput. Graph.}, 23(1):481--490, 2016. \href{https://doi.org/10.1109/TVCG.2016.2598609}
{doi: {{%
10\hspace{.1pt}\discretionary{.}{%
}{.}\hspace{.4pt}1109\discretionary{/}{%
}{/}TVCG\hspace{.1pt}\discretionary{.}{%
}{.}\hspace{.4pt}2016\hspace{.1pt}\discretionary{.}{%
}{.}\hspace{.4pt}2598609}}}


\bibitem{borkin2016beyond}
M.~A. Borkin, Z.~Bylinskii, N.~W. Kim, C.~M. Bainbridge, C.~S. Yeh, D.~Borkin, H.~Pfister, and A.~Oliva.
\newblock Beyond memorability: Visualization recognition and recall.
\newblock {\em IEEE Trans. Vis. Comput. Graph.}, 22(1):519--528, 2016. \href{https://doi.org/10.1109/TVCG.2015.2467732}
{doi: {{%
10\hspace{.1pt}\discretionary{.}{%
}{.}\hspace{.4pt}1109\discretionary{/}{%
}{/}TVCG\hspace{.1pt}\discretionary{.}{%
}{.}\hspace{.4pt}2015\hspace{.1pt}\discretionary{.}{%
}{.}\hspace{.4pt}2467732}}}


\bibitem{bostock2011d3}
M.~Bostock, V.~Ogievetsky, and J.~Heer.
\newblock D$^3$: Data-driven documents.
\newblock {\em IEEE Trans. Vis. Comput. Graph.}, 17(12):2301--2309, 2011. \href{https://doi.org/10.1109/TVCG.2011.185}
{doi: {{%
10\hspace{.1pt}\discretionary{.}{%
}{.}\hspace{.4pt}1109\discretionary{/}{%
}{/}TVCG\hspace{.1pt}\discretionary{.}{%
}{.}\hspace{.4pt}2011\hspace{.1pt}\discretionary{.}{%
}{.}\hspace{.4pt}185}}}


\bibitem{cairo2023art}
A.~Cairo.
\newblock {\em The art of insight: How great visualization designers think}.
\newblock John Wiley \& Sons, 2023.

\bibitem{chen2024visanatomy}
C.~Chen, H.~K. Bako, P.~Yu, J.~Hooker, J.~Joyal, S.~C. Wang, S.~Kim, J.~Wu, A.~Ding, L.~Sandeep, A.~Chen, C.~Sinha, and Z.~Liu.
\newblock {VisAnatomy}: An {SVG} chart corpus with fine-grained semantic labels.
\newblock {\em arXiv:2410.12268}, 2024.

\bibitem{chen2024Mystique}
C.~Chen, B.~Lee, Y.~Wang, Y.~Chang, and Z.~Liu.
\newblock Mystique: Deconstructing {SVG} charts for layout reuse.
\newblock {\em IEEE Trans. Vis. Comput. Graph.}, 30(1):447--457, 2024. \href{https://doi.org/10.1109/TVCG.2023.3327354}
{doi: {{%
10\hspace{.1pt}\discretionary{.}{%
}{.}\hspace{.4pt}1109\discretionary{/}{%
}{/}TVCG\hspace{.1pt}\discretionary{.}{%
}{.}\hspace{.4pt}2023\hspace{.1pt}\discretionary{.}{%
}{.}\hspace{.4pt}3327354}}}


\bibitem{chen2024beyond}
Q.~Chen, W.~Shuai, J.~Zhang, Z.~Sun, and N.~Cao.
\newblock Beyond numbers: Creating analogies to enhance data comprehension and communication with generative {AI}.
\newblock In {\em Proc. CHI},  article no. 377,  14 pages. ACM, 2024. \href{https://doi.org/10.1145/3613904.3642480}
{doi: {{%
10\hspace{.1pt}\discretionary{.}{%
}{.}\hspace{.4pt}1145\discretionary{/}{%
}{/}3613904\hspace{.1pt}\discretionary{.}{%
}{.}\hspace{.4pt}3642480}}}


\bibitem{cheng2024biscuit}
R.~Cheng, T.~Barik, A.~Leung, F.~Hohman, and J.~Nichols.
\newblock {BISCUIT}: Scaffolding {LLM}-generated code with ephemeral {UI}s in computational notebooks.
\newblock In {\em Proc. VL/HCC}, pp. 13--23. IEEE, 2024. \href{https://doi.org/10.1109/VL/HCC60511.2024.00012}
{doi: {{%
10\hspace{.1pt}\discretionary{.}{%
}{.}\hspace{.4pt}1109\discretionary{/}{%
}{/}VL\discretionary{/}{%
}{/}HCC60511\hspace{.1pt}\discretionary{.}{%
}{.}\hspace{.4pt}2024\hspace{.1pt}\discretionary{.}{%
}{.}\hspace{.4pt}00012}}}


\bibitem{coelho2020infomages}
D.~Coelho and K.~Mueller.
\newblock Infomages: Embedding data into thematic images.
\newblock {\em Comput. Graph. Forum}, 39(3):593--606, 2020. \href{https://doi.org/10.1111/cgf.14004}
{doi: {{%
10\hspace{.1pt}\discretionary{.}{%
}{.}\hspace{.4pt}1111\discretionary{/}{%
}{/}cgf\hspace{.1pt}\discretionary{.}{%
}{.}\hspace{.4pt}14004}}}


\bibitem{cui2021mixed}
W.~Cui, J.~Wang, H.~Huang, Y.~Wang, C.-Y. Lin, H.~Zhang, and D.~Zhang.
\newblock A mixed-initiative approach to reusing infographic charts.
\newblock {\em IEEE Trans. Vis. Comput. Graph.}, 28(1):173--183, 2021. \href{https://doi.org/10.1109/TVCG.2021.3114856}
{doi: {{%
10\hspace{.1pt}\discretionary{.}{%
}{.}\hspace{.4pt}1109\discretionary{/}{%
}{/}TVCG\hspace{.1pt}\discretionary{.}{%
}{.}\hspace{.4pt}2021\hspace{.1pt}\discretionary{.}{%
}{.}\hspace{.4pt}3114856}}}


\bibitem{omer2024pictorialattributes}
O.~Dahary, M.~Lu, O.~Patashnik, and D.~Cohen-Or.
\newblock {PictorialAttributes}: Depicting multiple attributes with realistic imaging.
\newblock In {\em ACM SIGGRAPH 2024 Posters},  article no. 69,  2 pages. ACM, 2024. \href{https://doi.org/10.1145/3641234.3671072}
{doi: {{%
10\hspace{.1pt}\discretionary{.}{%
}{.}\hspace{.4pt}1145\discretionary{/}{%
}{/}3641234\hspace{.1pt}\discretionary{.}{%
}{.}\hspace{.4pt}3671072}}}


\bibitem{di2023doom}
S.~Di~Bartolomeo, V.~Schetinger, J.~L. Adams, A.~M. McNutt, M.~El-Assady, and M.~Miller.
\newblock Doom or deliciousness: Challenges and opportunities for visualization in the age of generative models.
\newblock {\em Comput. Graph. Forum}, 42(3):423--435, 2023. \href{https://doi.org/10.1111/cgf.14841}
{doi: {{%
10\hspace{.1pt}\discretionary{.}{%
}{.}\hspace{.4pt}1111\discretionary{/}{%
}{/}cgf\hspace{.1pt}\discretionary{.}{%
}{.}\hspace{.4pt}14841}}}


\bibitem{dou2024Hierarchically}
S.~Dou, X.~Jiang, L.~Liu, L.~Ying, C.~Shan, Y.~Shen, X.~Dong, Y.~Wang, D.~Li, and C.~Zhao.
\newblock Hierarchically recognizing vector graphics and a new chart-based vector graphics dataset.
\newblock {\em IEEE Trans. Pattern Anal. Mach. Intell.}, 46(12):7556--7573, 2024. \href{https://doi.org/10.1109/TPAMI.2024.3394298}
{doi: {{%
10\hspace{.1pt}\discretionary{.}{%
}{.}\hspace{.4pt}1109\discretionary{/}{%
}{/}TPAMI\hspace{.1pt}\discretionary{.}{%
}{.}\hspace{.4pt}2024\hspace{.1pt}\discretionary{.}{%
}{.}\hspace{.4pt}3394298}}}


\bibitem{ge2025autopresent}
J.~Ge, Z.~Z. Wang, X.~Zhou, Y.-H. Peng, S.~Subramanian, Q.~Tan, M.~Sap, A.~Suhr, D.~Fried, G.~Neubig, and T.~Darrell.
\newblock {AutoPresent}: Designing structured visuals from scratch.
\newblock In {\em Proc. CVPR}, 2025.
\newblock arXiv:2501.00912.

\bibitem{goswami2025plotedit}
K.~Goswami, P.~Mathur, R.~Rossi, and F.~Dernoncourt.
\newblock {PlotEdit}: Natural language-driven accessible chart editing in {PDF}s via multimodal {LLM} agents.
\newblock In {\em Proc. ECIR}, pp. 130--134. Springer, 2025. \href{https://doi.org/10.1007/978-3-031-88720-8_22}
{doi: {{%
10\hspace{.1pt}\discretionary{.}{%
}{.}\hspace{.4pt}1007\discretionary{/}{%
}{/}978\discretionary{%
}{-}{-}3\discretionary{%
}{-}{-}031\discretionary{%
}{-}{-}88720\discretionary{%
}{-}{-}8\_22}}}


\bibitem{harper2014deconstructing}
J.~Harper and M.~Agrawala.
\newblock Deconstructing and restyling {D3} visualizations.
\newblock In {\em Proc. UIST}, pp. 253--262. ACM, 2014. \href{https://doi.org/10.1145/2642918.2647411}
{doi: {{%
10\hspace{.1pt}\discretionary{.}{%
}{.}\hspace{.4pt}1145\discretionary{/}{%
}{/}2642918\hspace{.1pt}\discretionary{.}{%
}{.}\hspace{.4pt}2647411}}}


\bibitem{harper2018Converting}
J.~Harper and M.~Agrawala.
\newblock Converting basic {D3} charts into reusable style templates.
\newblock {\em IEEE Trans. Vis. Comput. Graph.}, 24(3):1274--1286, 2018. \href{https://doi.org/10.1109/TVCG.2017.2659744}
{doi: {{%
10\hspace{.1pt}\discretionary{.}{%
}{.}\hspace{.4pt}1109\discretionary{/}{%
}{/}TVCG\hspace{.1pt}\discretionary{.}{%
}{.}\hspace{.4pt}2017\hspace{.1pt}\discretionary{.}{%
}{.}\hspace{.4pt}2659744}}}


\bibitem{harrison2015infographic}
L.~Harrison, K.~Reinecke, and R.~Chang.
\newblock Infographic aesthetics: Designing for the first impression.
\newblock In {\em Proc. CHI}, pp. 1187--1190. ACM, 2015. \href{https://doi.org/10.1145/2702123.2702545}
{doi: {{%
10\hspace{.1pt}\discretionary{.}{%
}{.}\hspace{.4pt}1145\discretionary{/}{%
}{/}2702123\hspace{.1pt}\discretionary{.}{%
}{.}\hspace{.4pt}2702545}}}


\bibitem{he2025Leveraging}
Y.~He, K.~Xu, S.~Cao, Y.~Shi, Q.~Chen, and N.~Cao.
\newblock Leveraging foundation models for crafting narrative visualization: A survey.
\newblock {\em IEEE Trans. Vis. Comput. Graph.}, 2025.
\newblock Early Access. \href{https://doi.org/10.1109/TVCG.2025.3542504}
{doi: {{%
10\hspace{.1pt}\discretionary{.}{%
}{.}\hspace{.4pt}1109\discretionary{/}{%
}{/}TVCG\hspace{.1pt}\discretionary{.}{%
}{.}\hspace{.4pt}2025\hspace{.1pt}\discretionary{.}{%
}{.}\hspace{.4pt}3542504}}}


\bibitem{jung2017chartsense}
D.~Jung, W.~Kim, H.~Song, J.-i. Hwang, B.~Lee, B.~Kim, and J.~Seo.
\newblock {ChartSense}: Interactive data extraction from chart images.
\newblock In {\em Proc. CHI}, pp. 6706--6717. ACM, 2017. \href{https://doi.org/10.1145/3025453.3025957}
{doi: {{%
10\hspace{.1pt}\discretionary{.}{%
}{.}\hspace{.4pt}1145\discretionary{/}{%
}{/}3025453\hspace{.1pt}\discretionary{.}{%
}{.}\hspace{.4pt}3025957}}}


\bibitem{kembhavi2016diagram}
A.~Kembhavi, M.~Salvato, E.~Kolve, M.~Seo, H.~Hajishirzi, and A.~Farhadi.
\newblock A diagram is worth a dozen images.
\newblock In {\em Proc. ECCV}, pp. 235--251. Springer, 2016. \href{https://doi.org/10.1007/978-3-319-46493-0_15}
{doi: {{%
10\hspace{.1pt}\discretionary{.}{%
}{.}\hspace{.4pt}1007\discretionary{/}{%
}{/}978\discretionary{%
}{-}{-}3\discretionary{%
}{-}{-}319\discretionary{%
}{-}{-}46493\discretionary{%
}{-}{-}0\_15}}}


\bibitem{kim2016data}
N.~W. Kim, E.~Schweickart, Z.~Liu, M.~Dontcheva, W.~Li, J.~Popovic, and H.~Pfister.
\newblock {Data-Driven Guides}: Supporting expressive design for information graphics.
\newblock {\em IEEE Trans. Vis. Comput. Graph.}, 23(1):491--500, 2016. \href{https://doi.org/10.1109/TVCG.2016.2598620}
{doi: {{%
10\hspace{.1pt}\discretionary{.}{%
}{.}\hspace{.4pt}1109\discretionary{/}{%
}{/}TVCG\hspace{.1pt}\discretionary{.}{%
}{.}\hspace{.4pt}2016\hspace{.1pt}\discretionary{.}{%
}{.}\hspace{.4pt}2598620}}}


\bibitem{kim2022stylette}
T.~S. Kim, D.~Choi, Y.~Choi, and J.~Kim.
\newblock Stylette: Styling the web with natural language.
\newblock In {\em Proc. CHI},  article no. 5,  17 pages. ACM, 2022. \href{https://doi.org/10.1145/3491102.3501931}
{doi: {{%
10\hspace{.1pt}\discretionary{.}{%
}{.}\hspace{.4pt}1145\discretionary{/}{%
}{/}3491102\hspace{.1pt}\discretionary{.}{%
}{.}\hspace{.4pt}3501931}}}


\bibitem{kumar2011bricolage}
R.~Kumar, J.~O. Talton, S.~Ahmad, and S.~R. Klemmer.
\newblock Bricolage: Example-based retargeting for web design.
\newblock In {\em Proc. CHI}, pp. 2197--2206. ACM, 2011. \href{https://doi.org/10.1145/1978942.1979262}
{doi: {{%
10\hspace{.1pt}\discretionary{.}{%
}{.}\hspace{.4pt}1145\discretionary{/}{%
}{/}1978942\hspace{.1pt}\discretionary{.}{%
}{.}\hspace{.4pt}1979262}}}


\bibitem{lewis2018system}
J.~R. Lewis.
\newblock The system usability scale: Past, present, and future.
\newblock {\em Int. J. Hum.-Comput. Interact.}, 34(7):577--590, 2018. \href{https://doi.org/10.1080/10447318.2018.1455307}
{doi: {{%
10\hspace{.1pt}\discretionary{.}{%
}{.}\hspace{.4pt}1080\discretionary{/}{%
}{/}10447318\hspace{.1pt}\discretionary{.}{%
}{.}\hspace{.4pt}2018\hspace{.1pt}\discretionary{.}{%
}{.}\hspace{.4pt}1455307}}}


\bibitem{liu2018learning}
T.~F. Liu, M.~Craft, J.~Situ, E.~Yumer, R.~Mech, and R.~Kumar.
\newblock Learning design semantics for mobile apps.
\newblock In {\em Proc. UIST},  11 pages, pp. 569--579. ACM, 2018. \href{https://doi.org/10.1145/3242587.3242650}
{doi: {{%
10\hspace{.1pt}\discretionary{.}{%
}{.}\hspace{.4pt}1145\discretionary{/}{%
}{/}3242587\hspace{.1pt}\discretionary{.}{%
}{.}\hspace{.4pt}3242650}}}


\bibitem{liu2024logomotion}
V.~Liu, R.~H. Kazi, L.-Y. Wei, M.~Fisher, T.~Langlois, S.~Walker, and L.~Chilton.
\newblock Logomotion: Visually-grounded code synthesis for creating and editing animation.
\newblock In {\em Proc. CHI},  article no. 157,  16 pages. ACM, 2025. \href{https://doi.org/10.1145/3706598.3714155}
{doi: {{%
10\hspace{.1pt}\discretionary{.}{%
}{.}\hspace{.4pt}1145\discretionary{/}{%
}{/}3706598\hspace{.1pt}\discretionary{.}{%
}{.}\hspace{.4pt}3714155}}}


\bibitem{liu2025manipulable}
Z.~Liu, C.~Chen, and J.~Hooker.
\newblock Manipulable semantic components: a computational representation of data visualization scenes.
\newblock {\em IEEE Trans. Vis. Comput. Graph.}, 31(1):732--742, 2025. \href{https://doi.org/10.1109/TVCG.2024.3456296}
{doi: {{%
10\hspace{.1pt}\discretionary{.}{%
}{.}\hspace{.4pt}1109\discretionary{/}{%
}{/}TVCG\hspace{.1pt}\discretionary{.}{%
}{.}\hspace{.4pt}2024\hspace{.1pt}\discretionary{.}{%
}{.}\hspace{.4pt}3456296}}}


\bibitem{liu2018dataillustrator}
Z.~Liu, J.~Thompson, A.~Wilson, M.~Dontcheva, J.~Delorey, S.~Grigg, B.~Kerr, and J.~Stasko.
\newblock {Data Illustrator}: Augmenting vector design tools with lazy data binding for expressive visualization authoring.
\newblock In {\em Proc. CHI},  13 pages. ACM, 2018. \href{https://doi.org/10.1145/3173574.3173697}
{doi: {{%
10\hspace{.1pt}\discretionary{.}{%
}{.}\hspace{.4pt}1145\discretionary{/}{%
}{/}3173574\hspace{.1pt}\discretionary{.}{%
}{.}\hspace{.4pt}3173697}}}


\bibitem{lu2020vif}
M.~Lu, C.~Wang, J.~Lanir, N.~Zhao, H.~Pfister, D.~Cohen-Or, and H.~Huang.
\newblock Exploring visual information flows in infographics.
\newblock In {\em Proc. CHI},  article no. 136,  12 pages. ACM, 2020. \href{https://doi.org/10.1145/3313831.3376263}
{doi: {{%
10\hspace{.1pt}\discretionary{.}{%
}{.}\hspace{.4pt}1145\discretionary{/}{%
}{/}3313831\hspace{.1pt}\discretionary{.}{%
}{.}\hspace{.4pt}3376263}}}


\bibitem{lu2024Misty}
Y.~Lu, A.~Leung, A.~Swearngin, J.~Nichols, and T.~Barik.
\newblock Misty: {UI} prototyping through interactive conceptual blending.
\newblock In {\em Proc. CHI},  article no. 1108,  17 pages. ACM, 2025. \href{https://doi.org/10.1145/3706598.3713924}
{doi: {{%
10\hspace{.1pt}\discretionary{.}{%
}{.}\hspace{.4pt}1145\discretionary{/}{%
}{/}3706598\hspace{.1pt}\discretionary{.}{%
}{.}\hspace{.4pt}3713924}}}


\bibitem{luo2021chartocr}
J.~Luo, Z.~Li, J.~Wang, and C.-Y. Lin.
\newblock {ChartOCR}: Data extraction from charts images via a deep hybrid framework.
\newblock In {\em Proc. WACV}, pp. 1917--1925, 2021. \href{https://doi.org/10.1109/WACV48630.2021.00196}
{doi: {{%
10\hspace{.1pt}\discretionary{.}{%
}{.}\hspace{.4pt}1109\discretionary{/}{%
}{/}WACV48630\hspace{.1pt}\discretionary{.}{%
}{.}\hspace{.4pt}2021\hspace{.1pt}\discretionary{.}{%
}{.}\hspace{.4pt}00196}}}


\bibitem{lupi2016Dear}
G.~Lupi, S.~Posavec, and M.~Popova.
\newblock {\em Dear Data}.
\newblock Princeton Architectural Press, 2016.

\bibitem{madan2021parsing}
S.~Madan, Z.~Bylinskii, C.~Nobre, M.~Tancik, A.~Recasens, K.~Zhong, S.~Alsheikh, A.~Oliva, F.~Durand, and H.~Pfister.
\newblock Parsing and summarizing infographics with synthetically trained icon detection.
\newblock In {\em Proc. PacificVis}, pp. 31--40, 2021. \href{https://doi.org/10.1109/PacificVis52677.2021.00012}
{doi: {{%
10\hspace{.1pt}\discretionary{.}{%
}{.}\hspace{.4pt}1109\discretionary{/}{%
}{/}PacificVis52677\hspace{.1pt}\discretionary{.}{%
}{.}\hspace{.4pt}2021\hspace{.1pt}\discretionary{.}{%
}{.}\hspace{.4pt}00012}}}


\bibitem{masson2023ChartDetective}
D.~Masson, S.~Malacria, D.~Vogel, E.~Lank, and G.~Casiez.
\newblock {ChartDetective}: Easy and accurate interactive data extraction from complex vector charts.
\newblock In {\em Proc. CHI},  article no. 147,  17 pages. ACM, 2023. \href{https://doi.org/10.1145/3544548.3581113}
{doi: {{%
10\hspace{.1pt}\discretionary{.}{%
}{.}\hspace{.4pt}1145\discretionary{/}{%
}{/}3544548\hspace{.1pt}\discretionary{.}{%
}{.}\hspace{.4pt}3581113}}}


\bibitem{mcnutt2021integrated}
A.~M. McNutt and R.~Chugh.
\newblock Integrated visualization editing via parameterized declarative templates.
\newblock In {\em Proc. CHI},  article no. 17,  14 pages. ACM, 2021. \href{https://doi.org/10.1145/3411764.3445356}
{doi: {{%
10\hspace{.1pt}\discretionary{.}{%
}{.}\hspace{.4pt}1145\discretionary{/}{%
}{/}3411764\hspace{.1pt}\discretionary{.}{%
}{.}\hspace{.4pt}3445356}}}


\bibitem{mendez2016iVoLVER}
G.~G. Méndez, M.~A. Nacenta, and S.~Vandenheste.
\newblock {iVoLVER}: Interactive visual language for visualization extraction and reconstruction.
\newblock In {\em Proc. CHI}, pp. 4073--4085. ACM, 2016. \href{https://doi.org/10.1145/2858036.2858435}
{doi: {{%
10\hspace{.1pt}\discretionary{.}{%
}{.}\hspace{.4pt}1145\discretionary{/}{%
}{/}2858036\hspace{.1pt}\discretionary{.}{%
}{.}\hspace{.4pt}2858435}}}


\bibitem{offenwanger2024datagarden}
A.~Offenwanger, T.~Tsandilas, and F.~Chevalier.
\newblock {DataGarden}: Formalizing personal sketches into structured visualization templates.
\newblock {\em IEEE Trans. Vis. Comput. Graph.}, 31(1):1268--1278, 2025. \href{https://doi.org/10.1109/TVCG.2024.3456336}
{doi: {{%
10\hspace{.1pt}\discretionary{.}{%
}{.}\hspace{.4pt}1109\discretionary{/}{%
}{/}TVCG\hspace{.1pt}\discretionary{.}{%
}{.}\hspace{.4pt}2024\hspace{.1pt}\discretionary{.}{%
}{.}\hspace{.4pt}3456336}}}


\bibitem{parsons2022Understanding}
P.~Parsons.
\newblock Understanding data visualization design practice.
\newblock {\em IEEE Trans. Vis. Comput. Graph.}, 28(1):665--675, 2023. \href{https://doi.org/10.1109/TVCG.2021.3114959}
{doi: {{%
10\hspace{.1pt}\discretionary{.}{%
}{.}\hspace{.4pt}1109\discretionary{/}{%
}{/}TVCG\hspace{.1pt}\discretionary{.}{%
}{.}\hspace{.4pt}2021\hspace{.1pt}\discretionary{.}{%
}{.}\hspace{.4pt}3114959}}}


\bibitem{poco2017ReverseEngineering}
J.~Poco and J.~Heer.
\newblock Reverse‐engineering visualizations: Recovering visual encodings from chart images.
\newblock {\em Comput. Graph. Forum}, 36(3):353--363, 2017. \href{https://doi.org/10.1111/cgf.13193}
{doi: {{%
10\hspace{.1pt}\discretionary{.}{%
}{.}\hspace{.4pt}1111\discretionary{/}{%
}{/}cgf\hspace{.1pt}\discretionary{.}{%
}{.}\hspace{.4pt}13193}}}


\bibitem{pollock2024bluefish}
J.~Pollock, C.~Mei, G.~Huang, E.~Evans, D.~Jackson, and A.~Satyanarayan.
\newblock Bluefish: Composing diagrams with declarative relations.
\newblock In {\em Proc. UIST},  article no. 23,  21 pages, 2024. \href{https://doi.org/10.1145/3654777.3676465}
{doi: {{%
10\hspace{.1pt}\discretionary{.}{%
}{.}\hspace{.4pt}1145\discretionary{/}{%
}{/}3654777\hspace{.1pt}\discretionary{.}{%
}{.}\hspace{.4pt}3676465}}}


\bibitem{qi2025sketch2data}
A.~Qi, T.~Tsandilas, A.~Shamir, and A.~Bousseau.
\newblock {Sketch2Data}: Recovering data from hand-drawn infographics.
\newblock {\em Comput. Graph.},  article no. 104251,  14 pages, 2025. \href{https://doi.org/10.1016/j.cag.2025.104251}
{doi: {{%
10\hspace{.1pt}\discretionary{.}{%
}{.}\hspace{.4pt}1016\discretionary{/}{%
}{/}j\hspace{.1pt}\discretionary{.}{%
}{.}\hspace{.4pt}cag\hspace{.1pt}\discretionary{.}{%
}{.}\hspace{.4pt}2025\hspace{.1pt}\discretionary{.}{%
}{.}\hspace{.4pt}104251}}}


\bibitem{qian2020retrieve}
C.~Qian, S.~Sun, W.~Cui, J.-G. Lou, H.~Zhang, and D.~Zhang.
\newblock Retrieve-then-adapt: Example-based automatic generation for proportion-related infographics.
\newblock {\em IEEE Trans. Vis. Comput. Graph.}, 27(2):443--452, 2020. \href{https://doi.org/10.1109/TVCG.2020.3030448}
{doi: {{%
10\hspace{.1pt}\discretionary{.}{%
}{.}\hspace{.4pt}1109\discretionary{/}{%
}{/}TVCG\hspace{.1pt}\discretionary{.}{%
}{.}\hspace{.4pt}2020\hspace{.1pt}\discretionary{.}{%
}{.}\hspace{.4pt}3030448}}}


\bibitem{ren2019Charticulator}
D.~Ren, B.~Lee, and M.~Brehmer.
\newblock Charticulator: Interactive construction of bespoke chart layouts.
\newblock {\em IEEE Trans. Vis. Comput. Graph.}, 25(1):789--799, 2019. \href{https://doi.org/10.1109/TVCG.2018.2865158}
{doi: {{%
10\hspace{.1pt}\discretionary{.}{%
}{.}\hspace{.4pt}1109\discretionary{/}{%
}{/}TVCG\hspace{.1pt}\discretionary{.}{%
}{.}\hspace{.4pt}2018\hspace{.1pt}\discretionary{.}{%
}{.}\hspace{.4pt}2865158}}}


\bibitem{satyanarayan2019critical}
A.~Satyanarayan, B.~Lee, D.~Ren, J.~Heer, J.~Stasko, J.~Thompson, M.~Brehmer, and Z.~Liu.
\newblock Critical reflections on visualization authoring systems.
\newblock {\em IEEE Trans. Vis. Comput. Graph.}, 26(1):461--471, 2019. \href{https://doi.org/10.1109/TVCG.2019.2934281}
{doi: {{%
10\hspace{.1pt}\discretionary{.}{%
}{.}\hspace{.4pt}1109\discretionary{/}{%
}{/}TVCG\hspace{.1pt}\discretionary{.}{%
}{.}\hspace{.4pt}2019\hspace{.1pt}\discretionary{.}{%
}{.}\hspace{.4pt}2934281}}}


\bibitem{satyanarayan2016vega}
A.~Satyanarayan, D.~Moritz, K.~Wongsuphasawat, and J.~Heer.
\newblock Vega-lite: A grammar of interactive graphics.
\newblock {\em IEEE Trans. Vis. Comput. Graph.}, 23(1):341--350, 2016. \href{https://doi.org/10.1109/TVCG.2016.2599030}
{doi: {{%
10\hspace{.1pt}\discretionary{.}{%
}{.}\hspace{.4pt}1109\discretionary{/}{%
}{/}TVCG\hspace{.1pt}\discretionary{.}{%
}{.}\hspace{.4pt}2016\hspace{.1pt}\discretionary{.}{%
}{.}\hspace{.4pt}2599030}}}


\bibitem{savva2011revision}
M.~Savva, N.~Kong, A.~Chhajta, L.~Fei-Fei, M.~Agrawala, and J.~Heer.
\newblock {ReVision}: Automated classification, analysis and redesign of chart images.
\newblock In {\em Proc. UIST}, pp. 393--402, 2011. \href{https://doi.org/10.1145/2047196.2047247}
{doi: {{%
10\hspace{.1pt}\discretionary{.}{%
}{.}\hspace{.4pt}1145\discretionary{/}{%
}{/}2047196\hspace{.1pt}\discretionary{.}{%
}{.}\hspace{.4pt}2047247}}}


\bibitem{schottler2024constraint}
S.~Sch{\"o}ttler, J.~Dykes, J.~Wood, U.~Hinrichs, and B.~Bach.
\newblock Constraint-based breakpoints for responsive visualization design and development.
\newblock {\em IEEE Trans. Vis. Comput. Graph.}, 2024.
\newblock Early Access. \href{https://doi.org/10.1109/TVCG.2024.3410097}
{doi: {{%
10\hspace{.1pt}\discretionary{.}{%
}{.}\hspace{.4pt}1109\discretionary{/}{%
}{/}TVCG\hspace{.1pt}\discretionary{.}{%
}{.}\hspace{.4pt}2024\hspace{.1pt}\discretionary{.}{%
}{.}\hspace{.4pt}3410097}}}


\bibitem{shen2024data}
L.~Shen, H.~Li, Y.~Wang, T.~Luo, Y.~Luo, and H.~Qu.
\newblock {Data Playwright}: Authoring data videos with annotated narration.
\newblock {\em IEEE Trans. Vis. Comput. Graph.}, 2024.
\newblock Early Access. \href{https://doi.org/10.1109/TVCG.2024.3477926}
{doi: {{%
10\hspace{.1pt}\discretionary{.}{%
}{.}\hspace{.4pt}1109\discretionary{/}{%
}{/}TVCG\hspace{.1pt}\discretionary{.}{%
}{.}\hspace{.4pt}2024\hspace{.1pt}\discretionary{.}{%
}{.}\hspace{.4pt}3477926}}}


\bibitem{shi2024ChartMimic}
C.~Shi, C.~Yang, Y.~Liu, B.~Shui, J.~Wang, M.~Jing, L.~Xu, X.~Zhu, S.~Li, Y.~Zhang, G.~Liu, X.~Nie, D.~Cai, and Y.~Yang.
\newblock {ChartMimic}: Evaluating {LMM}'s cross-modal reasoning capability via chart-to-code generation.
\newblock In {\em Proc. ICLR},  57 pages, 2025.
\newblock arXiv:2406.09961.

\bibitem{shi2025brickify}
X.~Shi, Y.~Wang, R.~Rossi, and J.~Zhao.
\newblock Brickify: Enabling expressive design intent specification through direct manipulation on design tokens.
\newblock In {\em Proc. CHI},  article no. 424,  20 pages. ACM, 2025. \href{https://doi.org/10.1145/3706598.3714087}
{doi: {{%
10\hspace{.1pt}\discretionary{.}{%
}{.}\hspace{.4pt}1145\discretionary{/}{%
}{/}3706598\hspace{.1pt}\discretionary{.}{%
}{.}\hspace{.4pt}3714087}}}


\bibitem{shi2023understanding}
Y.~Shi, T.~Gao, X.~Jiao, and N.~Cao.
\newblock Understanding design collaboration between designers and artificial intelligence: A systematic literature review.
\newblock {\em Proceedings of the ACM on Human-Computer Interaction}, 7(CSCW2),  article no. 368,  35 pages, 2023. \href{https://doi.org/10.1145/3610217}
{doi: {{%
10\hspace{.1pt}\discretionary{.}{%
}{.}\hspace{.4pt}1145\discretionary{/}{%
}{/}3610217}}}


\bibitem{shi2022supporting}
Y.~Shi, P.~Liu, S.~Chen, M.~Sun, and N.~Cao.
\newblock Supporting expressive and faithful pictorial visualization design with visual style transfer.
\newblock {\em IEEE Trans. Vis. Comput. Graph.}, 29(1):236--246, 2022. \href{https://doi.org/10.1109/TVCG.2022.3209486}
{doi: {{%
10\hspace{.1pt}\discretionary{.}{%
}{.}\hspace{.4pt}1109\discretionary{/}{%
}{/}TVCG\hspace{.1pt}\discretionary{.}{%
}{.}\hspace{.4pt}2022\hspace{.1pt}\discretionary{.}{%
}{.}\hspace{.4pt}3209486}}}


\bibitem{shin2021multi}
H.~V. Shin, J.~Warner, B.~Hartmann, C.~Gomes, H.~Winnemoeller, and W.~Li.
\newblock Multi-level correspondence via graph kernels for editing vector graphics designs.
\newblock In {\em Proc. GI}, pp. 97--107. CHCSS/SCDHM, 2021. \href{https://doi.org/10.20380/GI2021.12}
{doi: {{%
10\hspace{.1pt}\discretionary{.}{%
}{.}\hspace{.4pt}20380\discretionary{/}{%
}{/}GI2021\hspace{.1pt}\discretionary{.}{%
}{.}\hspace{.4pt}12}}}


\bibitem{snyder2025challenges}
L.~S. Snyder, C.~Wang, and S.~Drucker.
\newblock Challenges \& opportunities with llm-assisted visualization retargeting.
\newblock {\em ArXiv Preprint},  5 pages, 2025. \href{https://doi.org/10.48550/arXiv.2507.01436}
{doi: {{%
10\hspace{.1pt}\discretionary{.}{%
}{.}\hspace{.4pt}48550\discretionary{/}{%
}{/}arXiv\hspace{.1pt}\discretionary{.}{%
}{.}\hspace{.4pt}2507\hspace{.1pt}\discretionary{.}{%
}{.}\hspace{.4pt}01436}}}


\bibitem{iib}
D.~V. Society.
\newblock Information is beautiful awards.
\newblock \url{https://www.informationisbeautifulawards.com/about}, 2025.
\newblock Accessed on March 25, 2025.

\bibitem{song2024gvvst}
S.~Song, Y.~Zhang, Y.~Lin, H.~Qu, C.~Wang, and C.~Li.
\newblock {GVVST}: Image-driven style extraction from graph visualizations for visual style transfer.
\newblock {\em IEEE Trans. Vis. Comput. Graph.}, 2024.
\newblock Early Access. \href{https://doi.org/10.1109/TVCG.2024.3485701}
{doi: {{%
10\hspace{.1pt}\discretionary{.}{%
}{.}\hspace{.4pt}1109\discretionary{/}{%
}{/}TVCG\hspace{.1pt}\discretionary{.}{%
}{.}\hspace{.4pt}2024\hspace{.1pt}\discretionary{.}{%
}{.}\hspace{.4pt}3485701}}}


\bibitem{stefaner2025crisis}
M.~Stefaner.
\newblock Crisis? {What} crisis?---{Why} 2024 was a dead year for indie dataviz -- and how we’ll do much better in 2025.
\newblock \url{https://truth-and-beauty.net/texts/crisis-what-crisis}.
\newblock Accessed on March 25, 2025.

\bibitem{tian2024chartgpt}
Y.~Tian, W.~Cui, D.~Deng, X.~Yi, Y.~Yang, H.~Zhang, and Y.~Wu.
\newblock {ChartGPT}: Leveraging {LLM}s to generate charts from abstract natural language.
\newblock {\em IEEE Trans. Vis. Comput. Graph.}, 31(3):1731--1745, 2025. \href{https://doi.org/10.1109/TVCG.2024.3368621}
{doi: {{%
10\hspace{.1pt}\discretionary{.}{%
}{.}\hspace{.4pt}1109\discretionary{/}{%
}{/}TVCG\hspace{.1pt}\discretionary{.}{%
}{.}\hspace{.4pt}2024\hspace{.1pt}\discretionary{.}{%
}{.}\hspace{.4pt}3368621}}}


\bibitem{tsandilas2020structgraphics}
T.~Tsandilas.
\newblock {StructGraphics}: Flexible visualization design through data-agnostic and reusable graphical structures.
\newblock {\em IEEE Trans. Vis. Comput. Graph.}, 27(2):315--325, 2020. \href{https://doi.org/10.1109/TVCG.2020.3030476}
{doi: {{%
10\hspace{.1pt}\discretionary{.}{%
}{.}\hspace{.4pt}1109\discretionary{/}{%
}{/}TVCG\hspace{.1pt}\discretionary{.}{%
}{.}\hspace{.4pt}2020\hspace{.1pt}\discretionary{.}{%
}{.}\hspace{.4pt}3030476}}}


\bibitem{vaithilingam2024dynavis}
P.~Vaithilingam, E.~L. Glassman, J.~P. Inala, and C.~Wang.
\newblock {DynaVis}: Dynamically synthesized {UI} widgets for visualization editing.
\newblock In {\em Proc. CHI},  article no. 985,  17 pages. ACM, 2024. \href{https://doi.org/10.1145/3613904.3642639}
{doi: {{%
10\hspace{.1pt}\discretionary{.}{%
}{.}\hspace{.4pt}1145\discretionary{/}{%
}{/}3613904\hspace{.1pt}\discretionary{.}{%
}{.}\hspace{.4pt}3642639}}}


\bibitem{wang2025DataFormulator2}
C.~Wang, B.~Lee, S.~Drucker, D.~Marshall, and J.~Gao.
\newblock {Data Formulator 2:} iterative creation of data visualizations, with {AI} transforming data along the way.
\newblock In {\em Proc. CHI},  article no. 677,  17 pages. ACM, 2025. \href{https://doi.org/10.1145/3706598.3713296}
{doi: {{%
10\hspace{.1pt}\discretionary{.}{%
}{.}\hspace{.4pt}1145\discretionary{/}{%
}{/}3706598\hspace{.1pt}\discretionary{.}{%
}{.}\hspace{.4pt}3713296}}}


\bibitem{warner2023interactive}
J.~Warner, K.~W. Kim, and B.~Hartmann.
\newblock Interactive flexible style transfer for vector graphics.
\newblock In {\em Proc. UIST},  article no. 49,  14 pages. ACM, 2023. \href{https://doi.org/10.1145/3586183.3606751}
{doi: {{%
10\hspace{.1pt}\discretionary{.}{%
}{.}\hspace{.4pt}1145\discretionary{/}{%
}{/}3586183\hspace{.1pt}\discretionary{.}{%
}{.}\hspace{.4pt}3606751}}}


\bibitem{wilkinson2011grammar}
L.~Wilkinson.
\newblock The grammar of graphics.
\newblock In {\em Handbook of computational statistics: Concepts and methods}, pp. 375--414. Springer, 2011. \href{https://doi.org/10.1007/0-387-28695-0}
{doi: {{%
10\hspace{.1pt}\discretionary{.}{%
}{.}\hspace{.4pt}1007\discretionary{/}{%
}{/}0\discretionary{%
}{-}{-}387\discretionary{%
}{-}{-}28695\discretionary{%
}{-}{-}0}}}


\bibitem{wu2021ai4vis}
A.~Wu, Y.~Wang, X.~Shu, D.~Moritz, W.~Cui, H.~Zhang, D.~Zhang, and H.~Qu.
\newblock {AI4VIS}: Survey on artificial intelligence approaches for data visualization.
\newblock {\em IEEE Trans. Vis. Comput. Graph.}, 28(12):5049--5070, 2022. \href{https://doi.org/10.1109/TVCG.2021.3099002}
{doi: {{%
10\hspace{.1pt}\discretionary{.}{%
}{.}\hspace{.4pt}1109\discretionary{/}{%
}{/}TVCG\hspace{.1pt}\discretionary{.}{%
}{.}\hspace{.4pt}2021\hspace{.1pt}\discretionary{.}{%
}{.}\hspace{.4pt}3099002}}}


\bibitem{wu2024chat2svg}
R.~Wu, W.~Su, and J.~Liao.
\newblock {Chat2SVG}: Vector graphics generation with large language models and image diffusion models.
\newblock In {\em Proc. CVPR}, 2025.
\newblock arXiv:2411.16602.

\bibitem{wu2025what}
S.~Wu.
\newblock What killed innovation?
\newblock \url{https://www.shirleywu.studio/notebook/2025-02-innovation-killer}.
\newblock Accessed on March 25, 2025.

\bibitem{xia2018dataink}
H.~Xia, N.~Henry~Riche, F.~Chevalier, B.~De~Araujo, and D.~Wigdor.
\newblock {DataInk}: Direct and creative data-oriented drawing.
\newblock In {\em Proc. CHI},  article no. 223,  13 pages. ACM, 2018. \href{https://doi.org/10.1145/3173574.3173797}
{doi: {{%
10\hspace{.1pt}\discretionary{.}{%
}{.}\hspace{.4pt}1145\discretionary{/}{%
}{/}3173574\hspace{.1pt}\discretionary{.}{%
}{.}\hspace{.4pt}3173797}}}


\bibitem{xiao2023wytiwyr}
S.~Xiao, Y.~Hou, C.~Jin, and W.~Zeng.
\newblock {WYTIWYR}: A user intent-aware framework with multi-modal inputs for visualization retrieval.
\newblock {\em Comput. Graph. Forum}, 42(3):311--322, 2023. \href{https://doi.org/10.1111/cgf.14832}
{doi: {{%
10\hspace{.1pt}\discretionary{.}{%
}{.}\hspace{.4pt}1111\discretionary{/}{%
}{/}cgf\hspace{.1pt}\discretionary{.}{%
}{.}\hspace{.4pt}14832}}}


\bibitem{xiao2023spark}
S.~Xiao, S.~Huang, Y.~Lin, Y.~Ye, and W.~Zeng.
\newblock Let the chart spark: Embedding semantic context into chart with text-to-image generative model.
\newblock {\em IEEE Trans. Vis. Comput. Graph.}, 30(1):284--294, 2023. \href{https://doi.org/10.1109/TVCG.2023.3326913}
{doi: {{%
10\hspace{.1pt}\discretionary{.}{%
}{.}\hspace{.4pt}1109\discretionary{/}{%
}{/}TVCG\hspace{.1pt}\discretionary{.}{%
}{.}\hspace{.4pt}2023\hspace{.1pt}\discretionary{.}{%
}{.}\hspace{.4pt}3326913}}}


\bibitem{xiao2024typedance}
S.~Xiao, L.~Wang, X.~Ma, and W.~Zeng.
\newblock {TypeDance}: Creating semantic typographic logos from image through personalized generation.
\newblock In {\em Proc. CHI},  article no. 175,  18 pages. ACM, 2024. \href{https://doi.org/10.1145/3613904.3642185}
{doi: {{%
10\hspace{.1pt}\discretionary{.}{%
}{.}\hspace{.4pt}1145\discretionary{/}{%
}{/}3613904\hspace{.1pt}\discretionary{.}{%
}{.}\hspace{.4pt}3642185}}}


\bibitem{xie2024waitgpt}
L.~Xie, C.~Zheng, H.~Xia, H.~Qu, and C.~Zhu-Tian.
\newblock {WaitGPT}: Monitoring and steering conversational {LLM} agent in data analysis with on-the-fly code visualization.
\newblock In {\em Proc. UIST},  article no. 52,  14 pages. ACM, 2024. \href{https://doi.org/10.1145/3654777.3676374}
{doi: {{%
10\hspace{.1pt}\discretionary{.}{%
}{.}\hspace{.4pt}1145\discretionary{/}{%
}{/}3654777\hspace{.1pt}\discretionary{.}{%
}{.}\hspace{.4pt}3676374}}}


\bibitem{xie2023wakey}
L.~Xie, Z.~Zhou, K.~Yu, Y.~Wang, H.~Qu, and S.~Chen.
\newblock {Wakey-Wakey}: Animate text by mimicking characters in a {GIF}.
\newblock In {\em Proc. UIST},  article no. 98,  14 pages. ACM, 2023. \href{https://doi.org/10.1145/3586183.3606813}
{doi: {{%
10\hspace{.1pt}\discretionary{.}{%
}{.}\hspace{.4pt}1145\discretionary{/}{%
}{/}3586183\hspace{.1pt}\discretionary{.}{%
}{.}\hspace{.4pt}3606813}}}


\bibitem{ying2022MetaGlyph}
L.~Ying, X.~Shu, D.~Deng, Y.~Yang, T.~Tang, L.~Yu, and Y.~Wu.
\newblock {MetaGlyph}: Automatic generation of metaphoric glyph-based visualization.
\newblock {\em IEEE Trans. Vis. Comput. Graph.}, 29(1):331--341, 2022. \href{https://doi.org/10.1109/TVCG.2022.3209447}
{doi: {{%
10\hspace{.1pt}\discretionary{.}{%
}{.}\hspace{.4pt}1109\discretionary{/}{%
}{/}TVCG\hspace{.1pt}\discretionary{.}{%
}{.}\hspace{.4pt}2022\hspace{.1pt}\discretionary{.}{%
}{.}\hspace{.4pt}3209447}}}


\bibitem{yuan2021deep}
L.-P. Yuan, W.~Zeng, S.~Fu, Z.~Zeng, H.~Li, C.-W. Fu, and H.~Qu.
\newblock Deep colormap extraction from visualizations.
\newblock {\em IEEE Trans. Vis. Comput. Graph.}, 28(12):4048--4060, 2021. \href{https://doi.org/10.1109/TVCG.2021.3070876}
{doi: {{%
10\hspace{.1pt}\discretionary{.}{%
}{.}\hspace{.4pt}1109\discretionary{/}{%
}{/}TVCG\hspace{.1pt}\discretionary{.}{%
}{.}\hspace{.4pt}2021\hspace{.1pt}\discretionary{.}{%
}{.}\hspace{.4pt}3070876}}}


\bibitem{zhang2020DataQuilt}
J.~E. Zhang, N.~Sultanum, A.~Bezerianos, and F.~Chevalier.
\newblock {DataQuilt}: Extracting visual elements from images to craft pictorial visualizations.
\newblock In {\em Proc. CHI},  article no. 45,  13 pages. {ACM}, 2020. \href{https://doi.org/10.1145/3313831.3376172}
{doi: {{%
10\hspace{.1pt}\discretionary{.}{%
}{.}\hspace{.4pt}1145\discretionary{/}{%
}{/}3313831\hspace{.1pt}\discretionary{.}{%
}{.}\hspace{.4pt}3376172}}}


\bibitem{zhang2023editing}
S.~Zhang, J.~Ma, J.~Wu, D.~Ritchie, and M.~Agrawala.
\newblock Editing motion graphics video via motion vectorization and transformation.
\newblock {\em ACM Trans. Gr.}, 42(6),  article no. 229,  13 pages, 2023. \href{https://doi.org/10.1145/3618316}
{doi: {{%
10\hspace{.1pt}\discretionary{.}{%
}{.}\hspace{.4pt}1145\discretionary{/}{%
}{/}3618316}}}


\bibitem{zhang2023oldvisonline}
Y.~Zhang, R.~Jiang, L.~Xie, Y.~Zhao, C.~Liu, T.~Ding, S.~Chen, and X.~Yuan.
\newblock {OldVisOnline}: Curating a dataset of historical visualizations.
\newblock {\em IEEE Trans. Vis. Comput. Graph.}, 30(1):551--561, 2024. \href{https://doi.org/10.1109/TVCG.2023.3326908}
{doi: {{%
10\hspace{.1pt}\discretionary{.}{%
}{.}\hspace{.4pt}1109\discretionary{/}{%
}{/}TVCG\hspace{.1pt}\discretionary{.}{%
}{.}\hspace{.4pt}2023\hspace{.1pt}\discretionary{.}{%
}{.}\hspace{.4pt}3326908}}}


\bibitem{zhou2024DataPictorial}
T.~Zhou, G.~Y.-Y. Chan, S.~Guo, J.~Hoffswell, C.~Xiao, V.~S. Bursztyn, and E.~Koh.
\newblock {Data Pictorial}: Deconstructing raster images for data-aware animated vector posters.
\newblock In {\em Proc. UIST Adjunct},  article no. 98,  3 pages. ACM, 2024. \href{https://doi.org/10.1145/3672539.3686353}
{doi: {{%
10\hspace{.1pt}\discretionary{.}{%
}{.}\hspace{.4pt}1145\discretionary{/}{%
}{/}3672539\hspace{.1pt}\discretionary{.}{%
}{.}\hspace{.4pt}3686353}}}


\bibitem{zhou2024epigraphics}
T.~Zhou, J.~Huang, and G.~Y.-Y. Chan.
\newblock Epigraphics: Message-driven infographics authoring.
\newblock In {\em Proc. CHI},  18 pages. ACM, 2024. \href{https://doi.org/10.1145/3613904.3642172}
{doi: {{%
10\hspace{.1pt}\discretionary{.}{%
}{.}\hspace{.4pt}1145\discretionary{/}{%
}{/}3613904\hspace{.1pt}\discretionary{.}{%
}{.}\hspace{.4pt}3642172}}}


\bibitem{zhu2019towards}
C.~Zhu-Tian, Y.~Wang, Q.~Wang, Y.~Wang, and H.~Qu.
\newblock Towards automated infographic design: Deep learning-based auto-extraction of extensible timeline.
\newblock {\em IEEE Trans. Vis. Comput. Graph.}, 26(1):917--926, 2019. \href{https://doi.org/10.1109/TVCG.2019.2934810}
{doi: {{%
10\hspace{.1pt}\discretionary{.}{%
}{.}\hspace{.4pt}1109\discretionary{/}{%
}{/}TVCG\hspace{.1pt}\discretionary{.}{%
}{.}\hspace{.4pt}2019\hspace{.1pt}\discretionary{.}{%
}{.}\hspace{.4pt}2934810}}}


\end{thebibliography}
